\documentclass[a4paper,11pt]{article}
\usepackage{jheppub} 
\usepackage[section]{placeins}
\usepackage[T1]{fontenc} 
\usepackage{graphicx}
\graphicspath{ {figures/} }
\usepackage{amsmath}
\usepackage{bm}
\usepackage{subfigure}
\usepackage{amssymb}
\usepackage{mathrsfs}
\usepackage{braket}
\usepackage{comment}

\newcommand{\comments}[1]{}

\usepackage{slashed}
\usepackage{epsfig,psfrag,rotating,soul}
\usepackage{rotfloat}
\usepackage{empheq}  

\allowdisplaybreaks

\title{\boldmath  Quantumness of top quark pairs produced at LHC within SMEFT framework} 

\author[a,b]{Amir Subba }

\author[a,b]{Yu Shi }

\affiliation[a]{ Wilczek Quantum Center, Shanghai Institute for Advanced Studies, Shanghai 201315, China }  

\affiliation[b]{
University of Science and Technology of China, Hefei 230026, China }

\emailAdd{amirsubba@ustc.edu.cn}

\emailAdd{yu\_shi@ustc.edu.cn}  

\abstract{
Top and anti-top quark pair production at LHC provides a unique setting to probe non-classical correlations at the TeV scale. We study quantum information (QI) properties of the $t\bar{t}$ spin state in $pp$ collisions at $\sqrt{s}=13$ TeV within the Standard Model Effective Field Theory (SMEFT), focusing on dimension-6 operators that induce anomalous chromo- and weak dipole moments of the top quark within their current experimental bounds. The $t\bar{t}$ spin density matrix is reconstructed from the joint angular distribution of the final state charged leptons in the $k$-$r$-$n$ helicity basis. We analyze three complementary QI quantities, concurrence-based quantum entanglement (QE), geometric quantum discord (GQD), and the Bell parameter, across four $t\bar{t}$ invariant-mass bins. Within the Standard Model (SM), non-vanishing QE appears only near threshold ($m_{t\bar{t}}\lesssim 400$ GeV), while GQD remains nonzero across the full phase space, indicating persistent non-classical correlations even for separable states. Anomalous chromo-dipole interactions modify these observables primarily near threshold: $\hat{\mu}_t$ induces asymmetric shifts, whereas $\hat{d}_t$ produces a mild symmetric response without Bell inequality violation. Among weak dipole operators, the CP-even coupling $C_2^V$ generates the largest deformation of the QI observables, while $\Delta C_1^{A,V}$ leave them unchanged.  
These results demonstrate that QI observables derived from the $t\bar{t}$ spin density matrix provide a complementary probe of anomalous top-quark interactions with distinct sensitivity to CP-even and CP-odd operator structures.
}

\begin{document}

\maketitle

\section{Introduction}
\label{sec:intro}
The top quark, with a mass of $\sim173$~GeV~\cite{ParticleDataGroup:2022pth}, occupies a unique role in both collider physics and fundamental quantum mechanics. Its large mass implies a very short lifetime $(\sim 10^{-25}~s)$, even shorter than the timescale for hadronization $(\tau_{\mathrm{Had}}\sim 1/\Lambda_{\mathrm{QCD}}\sim 10^{-24}~s)$ and spin decorrelation $(\tau_{\mathrm{Decorr}}\sim hm_t/(2\pi \Lambda_{\mathrm{QCD}}^2)\sim 10^{-21}~s)$~\cite{Grossman:2008qh,Mahlon:2010gw}, allowing top quarks to decay before their quantum state is washed out by non-perturbative QCD effects. This property ensures that the spin information of the top quark at production is directly transferred to its decay products and can be experimentally accessed through angular distributions~\cite{Mahlon:1995zn,Stelzer:1995gc}. 

In the  Large Hadron Collider (LHC) at CERN, top quarks are predominantly produced in pairs via gluon fusion. At QCD leading order, $t\bar{t}$ pairs are unpolarized: longitudinal polarization vanishes due to parity conservation, while transverse polarization is suppressed by time-reversal invariance. In SM, small residual polarizations arise beyond leading order. Electroweak corrections induce a small longitudinal component, and absorptive parts of one-loop QCD amplitudes generate a transverse polarization at the percent level~\cite{Bernreuther:2013aga,Bernreuther:2015yna}.

The first experimental evidence for $t\bar{t}$ spin correlations was reported by D0~\cite{D0:2011kcb}, with a significance exceeding $3\sigma$. This was followed by measurements from ATLAS~\cite{ATLAS:2012ao}, which excluded the zero spin-correlation hypothesis at $5.1\sigma$, and subsequent results from CMS~\cite{CMS:2013roq,CMS:2015cal,CMS:2016piu,CMS:2019kzp,CMS:2019nrx} and ATLAS~\cite{ATLAS:2014abv,ATLAS:2014aus,ATLAS:2019zrq} across multiple final states and center-of-mass energies. These measurements firmly establish the presence of spin correlations in $t\bar{t}$ production. Accordingly, $t\bar{t}$ pairs produced at high-energy colliders can be treated as high-energy physical realizations of two-qubit systems with experimentally accessible spin degrees of freedom, providing a unique platform to investigate quantum correlations at the electroweak scale.

Quantum correlations describe statistical dependencies between subsystems that cannot be reproduced by classical joint probability distributions. The most prominent manifestation is quantum entanglement (QE)~\cite{shiwu,shiroad,shihistoric}, for which the composite density matrix cannot be factorized into a convex combination of product density matrices. More general measures, such as quantum discord~\cite{Ollivier:2001fdq} (QD), quantify non-classical correlations that may persist even in separable mixed states through the non-zero difference between total and classically accessible correlations under local measurements. These concepts characterize the structure of quantum information shared between subsystems and extend beyond entanglement alone. Quantum discord is a useful quantum information resource with its utilization in quantum computation~\cite{Datta:2007err,Wang:2019qwc} and quantum cryptography~\cite{Pirandola:2014uqc,Dakic:2012ogw}.  More recently such quantum-information resources have begun to attract attention in high-energy contexts~\cite{Han:2024ugl,Afik:2022dgh}. However, due to the minimization procedure involved in the QD, the exact analytic closed form expression is only available for a two qubit system~\cite{Luo:2008ecu}. Alternative forms of QD are suggested~\cite{Bera:2017lmd}.

With  large data samples accumulated at  LHC, it has become possible to probe such quantum features experimentally in $t\bar{t}$ production. Measurements of polarization and spin correlations constrain the joint spin density matrix and thereby enable the construction of QI observables. An example is the entanglement marker $D=\mathrm{Tr}[C]/$~\cite{Afik:2020onf}, proposed as a criterion based on the spin–correlation matrix. In the threshold region, CMS~\cite{CMS:2024pts} has measured $D=-0.480^{+0.026}_{-0.029}$ and ATLAS~\cite{ATLAS:2023fsd} has reported $D=-0.537\pm0.002,(\mathrm{stat.})\pm0.019,(\mathrm{syst.})$. Both results satisfy the condition $D<-1/3$, providing experimental evidence for QE between the top and antitop quarks. Beyond entanglement, increasingly stringent notions of non-classicality—such as quantum discord, quantum steering~\cite{Schrodinger1936,Wiseman:2007hyt}, and Bell nonlocality—offer a hierarchical framework for characterizing quantum correlations in collider-produced systems.

On the theoretical front, a substantial body of work has investigated quantum entanglement (QE) and Bell nonlocality in high-energy processes.  Among many others, one of us has studied QE in pseudoscalar mesons~\cite{shiteleportation,shiteleportation2,huangshi,shicp,shicp2,huangshi2,shiyang,shiyangleggett,shiyangbell}, and in relativistic quantum field theory~\cite{shifield}. In recent years, the fermionic two-qubit structure  has been extensively explored in $t\bar{t}$ system at LHC~\cite{Afik:2020onf,Afik:2025grr,subba2024,Cheng:2024btk,Aguilar-Saavedra:2024vpd,Barr:2024djo,Aguilar-Saavedra:2024hwd,Maltoni:2024tul,Cheng:2023qmz,Aguilar-Saavedra:2023hss,Aguilar-Saavedra:2022uye,Aoude:2022imd,Afik:2022kwm,Severi:2021cnj,Fabbrichesi:2021npl,Han:2023fci} and  $e^-e^+$ colliders~\cite{Ehataht:2023zzt,subba2024,Han:2025ewp,Guo:2026yhz}.  Beyond foundational aspects, QI quantities in top-quark production have emerged as complementary probes of physics beyond the Standard Model (BSM)~\cite{Fabbrichesi:2022ovb,Maltoni:2024tul}. New interactions can modify the production spin density matrix, thereby altering the structure of spin correlations and the associated quantum correlations. In particular, CP-violating effects can induce characteristic asymmetries in the spin correlations  that propagate into quantum-information measures such as discord or steering~\cite{Afik:2022dgh}, providing conceptually distinct and potentially sensitive signatures of new physics.

In this paper, we systematically investigate QI quantities  in the $t\bar{t}$ systems, both within SM  and in the SMEFT framework. We consider dimension-6 operators modifying the $t\bar{t}Z$ and $gt\bar{t}$ vertices, which generate anomalous weak and chromo dipole moments of the top quark and thereby deform the spin density matrix of the produced pair through their effects on the neutral-current and gluonic production mechanisms. Three complementary observables are studied: concurrence as a measure  of QE, GQD as a probe of non-classical correlations beyond entanglement, and the 
Bell-CHSH parameter as a test of genuine nonlocality. Their dependence on the anomalous couplings is examined across four invariant-mass bins of the $t\bar{t}$ 
system, spanning the threshold through the boosted regime, thereby mapping out the 
interplay between kinematics and new-physics sensitivity. In addition, we demonstrate 
that the difference of quantum discords (QDs), $\Delta\mathcal{D}[\rho_{t\bar{t}}]$, 
constructed from measurements on the two subsystems, provides a direct and 
theoretically transparent probe of CP-violating interactions. Taken together, this 
analysis establishes QI quantities as sensitive and complementary 
tools for probing the anomalous top-quark sector at  LHC.

The rest of the paper is organized as follows. In Sec.~\ref{sec:formalism}, we discuss the reconstruction of the quantum state of the $t\bar{t}$ system from the joint angular distribution of the final decayed fermions in the helicity basis. In Sec.~\ref{sec:corr}, we examine the non-classicality of the $t\bar{t}$ system, focusing in particular on  concurrence, GQD and Bell violation. The SMEFT description of $t\bar{t}$ production is presented in Sec.~\ref{sec:smeft}. We describe the results  in Sec.~\ref{sec:results}. Finally, we conclude in Sec.~\ref{sec:con}.

\section{Reference basis, Angular Distribution and Density Matrix for the \texorpdfstring{$t\bar{t}$}{ttbar} System}
\label{sec:formalism}

We consider top quark pair production at   LHC, followed by the di-leptonic decay of the top quarks. 
The dominant partonic sub-processes are $q\bar{q} \to t\bar{t}$ and $gg \to t\bar{t}$, which do not interfere with each other. 
Using the narrow-width approximation for the top quark, the squared amplitude for the full process $t\bar{t} \to \ell^+ \nu b \,\ell^- \bar{\nu}\bar{b}$ factorizes into production and decay parts:
\begin{align}
    \label{eq:mat}
    |\mathcal{M}(q\bar{q}/gg \to t\bar{t} \to \ell^+ \nu b \,\ell^- \bar{\nu}\bar{b})|^2 
    \propto \mathrm{Tr}\Big[(D_t \otimes D_{\bar{t}}) \, R_{t\bar{t}}^{q\bar{q}/gg} \Big],
\end{align}
where $R_{t\bar{t}}$ is the production spin density matrix and $D'$s are the decay density matrices of the top quarks, respectively. 

The production spin density matrix can be decomposed in the top and anti-top spin spaces using a Pauli matrix basis as~\cite{Bernreuther:2015yna}
\begin{align}
    \label{eq:pdm}
    R_{t\bar{t}}^I &= f^I \Big[ 
    \widetilde{A}\, \mathbb{I}_4 
    + \sum_i \widetilde{B}_i^+ (\tau_i \otimes \mathbb{I}_2) 
    + \sum_i \widetilde{B}_i^- (\mathbb{I}_2 \otimes \tau_i) 
    + \sum_{i,j} \widetilde{C}_{ij}^{t\bar{t}} (\tau_i \otimes \tau_j) 
    \Big],
\end{align}
where $f^{gg} = (4\pi\alpha_s)^2/N_C(N_C-1)$ and $f^{q\bar{q}} = (N_C^2-1)(4\pi\alpha_s)^2/N_C^2$, with $N_C$ being the number of color degrees of freedom. 
Here, $A$ determines the total $t\bar{t}$ cross section and kinematics, $\widetilde{B}^\pm$ are vectors encoding the top/anti-top polarization (within SM, they vanish due to QCD discrete symmetry), and $\widetilde{C}_{ij}$ characterizes spin correlations between the two quarks. 
The most general spin density matrix for a pair of qubits then can be obtained from $R$ by proper normalization of Eq.~\eqref{eq:pdm} as~\cite{Afik:2020onf}
\begin{align}
    \rho_{t\bar{t}} = \frac{R_{t\bar{t}}}{\mathrm{Tr}[R_{t\bar{t}}]} = \frac{R}{4\widetilde{A}},\qquad B_i^\pm = \frac{\widetilde{B}^\pm_i}{\widetilde{A}},\qquad C_{ij} = \frac{\widetilde{C}_{ij}}{\widetilde{A} }.
\end{align}
With this normalization, the proper spin density matrix for $t\bar{t}$ system is given by
\begin{align}
    \label{eq:sdm}
    \rho_{t\bar{t}} &= 
     \frac{1}{4} \Big[ 
    \mathbb{I}_4 
    + \sum_i B_i^+ (\tau_i \otimes \mathbb{I}_2) 
    + \sum_i B_i^- (\mathbb{I}_2 \otimes \tau_i) 
    + \sum_{i,j} C_{ij}^{t\bar{t}} (\tau_i \otimes \tau_j) 
    \Big].
\end{align}
The above $t\bar{t}$ quantum state is a mixed state comprised of the density matrices of the quark annihilation and gluon fusion partonic production channels. For quark–antiquark annihilation, the spin density matrix takes the form~\cite{Han:2024ugl}
\begin{equation}
\label{eq:qq}
\rho_{q\bar q} = a\,\rho^{(+)} + (1-a)\,\rho_{\text{mix}}^{(X)},
\qquad
a = \frac{\beta^2}{2 - \beta^2},\,\qquad \beta = \sqrt{1-4m_t^2/m_{t\bar{t}}^2}.
\end{equation}
The state $\rho_{\text{mix}}^{(i)}$ describes a classical mixture of both spins aligned along the $i$-direction. Near production threshold ($\beta \to 0$), the state approaches the separable mixed state $\rho_{\text{mix}}^{(X)}$. In contrast, in the highly boosted regime ($\beta \to 1$), the state becomes the pure maximally entangled Bell state $\rho^{(+)}$. For gluon fusion, the spin density matrix is more involved and can be written as
\begin{equation}
\label{eq:gg}
\rho_{gg} = a_1 \rho^{(+)} + a_2 \rho^{(-)} 
+ a_3 \rho_{\text{mix}}^{(X)} 
+ a_4 \rho_{\text{mix}}^{(Y)},
\end{equation}
with
\begin{equation}
a_1 = \frac{\beta^4}{1 + 2\beta^2 - 2\beta^4}, \quad
a_2 = \frac{(1-\beta^2)^2}{1 + 2\beta^2 - 2\beta^4}, \quad
a_3 = a_4 = \frac{2\beta^2(1-\beta^2)}{1 + 2\beta^2 - 2\beta^4},
\end{equation}
subject to the normalization condition  $\sum_ia_i = 1$~\cite{Barr:2024djo}. In this case, the limiting behavior is reversed compared to the $q\bar q$ channel. Near threshold ($\beta \to 0$), the state approaches the pure entangled Bell state $\rho^{(-)}$, while in the ultra-relativistic limit ($\beta \to 1$), it approaches $\rho^{(+)}$. At intermediate values of $\beta$, the presence of non-zero $a_3$ and $a_4$ induces a substantial mixed component, reflecting the non-trivial interplay between different spin configurations in gluon-fusion production.

Moreover, for  a given quantum state of  $t\bar{t}$ as a pair of particles, any measure of quantumness, viz., entanglement, quantum discord, etc., is  an intrinsic property and does not depend on the choice of measurement basis, as long as the density matrix is reconstructed from an ensemble of identically prepared events sharing a common spin quantization axis. If instead an event-dependent axis is chosen, the reconstructed object represents an angular-averaged mixture of sub-states rather than a true quantum state. Such states are referred to as an ``fictitious state''~\cite{Afik:2022kwm,Cheng:2023qmz,Han:2023fci} and the measurement or observation of quantumness on such a system may not directly  correspond to quantumness in a genuine quantum state. For this reason, defining a uniform quantization axis across events is crucial for spin-related studies~\cite{Cheng:2024btk}.

In the $t\bar{t}$ center-of-mass frame, an orthonormal basis $\{\hat{k}, \hat{r}, \hat{n}\}$ is defined~\cite{Bernreuther:2015yna}. 
The helicity axis $\hat{k}$ is aligned along the top quark's direction of motion, $\hat{p}$ denotes the direction of the incoming parton, and $\Theta$ is the top-quark scattering angle. The component of the beam direction that is orthogonal to $\hat{k}$ is
\begin{equation}
    \hat{r} = \frac{\hat{p} - \hat{k}\cos\Theta}{\sin\Theta},
\end{equation}
while the remaining axis in the production plane is
\begin{equation}
    \hat{n} = \hat{r} \times \hat{k},
\end{equation}
The axes form a right-handed system. Additionally, for $gg$ initial states, Bose--Einstein symmetry requires a redefinition to avoid vanishing spin correlation coefficients:
\begin{equation}
    \{\hat{k}, \hat{r}, \hat{n}\} 
    \longrightarrow 
    \{\hat{k}, \mathrm{sign}(\cos\Theta)\,\hat{r}, \mathrm{sign}(\cos\Theta)\,\hat{n}\}
\end{equation}
i.e., we have used the sign of the cosine of the top quark
scattering angle, which is odd under Bose-Einstein symmetry, to define a ``forward'' direction for each event. The helicity basis, by contrast, is not a fixed basis because the axes change event-by-event. Performing the summation over many events does not measure a parameter of density matrix but rather its expectation value~\cite{Afik:2022kwm} since the basis is different event by event.

Experimentally, the spin density matrix is reconstructed from the joint angular distributions of the decayed final leptons in the respective parent top quarks rest frames. An alternative method~\cite{Cheng:2024rxi} also exists based on the usage of production kinematics to construct the density matrix. We stick to decay method, in the current analysis. Eq.~\eqref{eq:mat} can be expanded to obtain the normalized four-fold differential distribution of the final decayed leptons as
\begin{align}
\label{eq:jad}
\frac{1}{\sigma}\frac{d^{4}\sigma}{d\Omega_{\ell^+}d\Omega_{\ell^-}} 
&= \frac{1}{16\pi^2} \Big[ 
1 
+ \alpha_{\ell^+} B_i^+\cdot \hat{\ell}^+_i
+ \alpha_{\ell^-} B_i^- \cdot \hat{\ell}^- 
+ \alpha_{\ell^+} \alpha_{\ell^-} \hat{\ell}^+ \cdot C.\hat{\ell}^-
\Big],
\end{align}
where $\sigma$ is the $t\bar{t}$ production cross section and $d\Omega_{\ell^\pm}$ are the solid angle of charged leptons in their parent top quark and antiquark
rest frames, and $\hat{\ell}^\pm$ are the corresponding unit vectors in their mother rest frame. The spin-analyzing power $\alpha_{\ell^\pm}$ of a daughter charged leptons quantifies how effectively the spin state of the parent top quark can be reconstructed from the kinematics of that leptons. It appears in the differential decay distribution of the top quark:
\begin{align}
\frac{1}{\Gamma}\frac{d\Gamma}{d\cos\Theta} = \frac{1}{2}\Big(1 + \alpha_f\,B^+\, \cos\Theta\Big),
\end{align}
where $\Theta$ is the angle between the top quark’s polarization vector $B^+$ and the direction of the daughter fermion. Within  SM and at leading order, the spin-analyzing power takes the values $\alpha_{l,d} = +1$ for down-type fermions from $W$ decays, $\alpha_{\nu,u} = -0.3$ for up-type fermions, and $\alpha_b = -0.4$ for $b$ quarks~\cite{Brandenburg:2002xr}.

After integrating out the $\phi$ degree of freedom in Eq.~\eqref{eq:jad}, the differential rate becomes just a function of $\cos\theta$ as
\begin{align}
    \frac{1}{\sigma}\frac{d^2\sigma}{d\cos\theta_{\ell^+}^i d\cos\theta_{\ell^-}^-} &= \frac{1}{4}\Big[1+B_{\ell^+}^i\cos\theta_{\ell^+}^i+B_{\ell^-}^j\cos\theta_{\ell^-}^j +\alpha_{\ell^+}\alpha_{\ell^-}C_{ij}\cos\theta_{\ell^+}^i\cos\theta_{\ell^-}^j\Big],
\end{align}
where $\cos\theta_{\ell^\pm}^{i(j)} = \hat{\ell}^\pm_i\cdot \hat{i}$ is the cosine distribution of the  charged lepton, measured with respect to axis $i(j)$ in the rest frame of its parent top quark (anti top quark). Upon integrating w.r.t  $\cos\theta_{\ell^\pm},\, \cos\theta_{\ell^+}\cos\theta_{\ell^-}$ and change of variables, we obtain a single differential distribution~\cite{CMS:2019nrx},
\begin{align}
    &\frac{1}{\sigma}\frac{d\sigma}{d\cos\theta_{\ell^\pm}^i} = \frac{1}{2}\Big(1 + \alpha_{\ell^\pm} B_i^\pm\cos\theta_{\ell^\pm}^i\Big),\nonumber\\
    &\frac{1}{\sigma}\frac{d\sigma}{d\cos\theta_{\ell^+}d\cos\theta_{\ell^-}} = \frac{1}{2}(1 - C_{ij}\cos\theta_{\ell^+}^i\cos\theta_{\ell^-}^j)\,\mathrm{ln}\Big(\frac{1}{|\cos\theta_{\ell^+}^i\cos\theta_{\ell^-}^j|}\Big),\nonumber\\
    &\frac{1}{\sigma}\frac{d\sigma}{dx_{\pm}} = \frac{1}{2}(1-\frac{C_{ij}\pm C_{ji}}{2}x_\pm)\,\cos^-|x_\pm|,\, x_\pm = \cos\theta_{\ell^+}^i\cos\theta_{\ell^-}^j\pm \cos\theta_{\ell^+}^j\cos\theta_{\ell^-}^i,\, \mathrm{for}\,i\neq j.
\end{align}

\begin{table}[!htb]
    \centering
    \caption{Observables and their corresponding measured coefficients of production spin density matrix and parity and CP symmetry properties.}
    \label{tab:obs}
    \begin{tabular*}{\textwidth}{@{\extracolsep{\fill}} c c c @{}}
    \hline
       $f(\theta_{\ell^+},\theta_{\ell^-})$  & Coefficient & Symmetries  \\
         \hline
        $\cos\theta_{\ell^\pm}^k$ & $B_k^\pm$ &P-Odd, CP-Even \\
        $\cos\theta_{\ell^\pm}^r$ & $B_r^\pm$ &P-Odd, CP-Even \\
        $\cos\theta_{\ell^\pm}^n$ & $B_n^\pm$ &P-Even, CP-Even \\
        $\cos\theta_{\ell^+}^k \cos\theta_{\ell^-}^k$ & $C_{kk}$ & P-Even, CP-Even\\
        $\cos\theta_{\ell^+}^r \cos\theta_{\ell^-}^r$ & $C_{rr}$ & P-Even, CP-Even\\
        $\cos\theta_{\ell^+}^n \cos\theta_{\ell^-}^n$ & $C_{nn}$ & P-Even, CP-Even\\
        $\cos\theta_{\ell^+}^r\cos\theta_{\ell^-}^k+\cos\theta_{\ell^+}^k\cos\theta_{\ell^-}^r$ & $C_{rk} + C_{kr}$ & P-Even, CP-Even\\
        $\cos\theta_{\ell^+}^r\cos\theta_{\ell^-}^k-\cos\theta_{\ell^+}^k\cos\theta_{\ell^-}^r$ & $C_{rk} - C_{kr}$ & P-Even, CP-Odd\\
        $\cos\theta_{\ell^+}^n\cos\theta_{\ell^-}^r+\cos\theta_{\ell^+}^r\cos\theta_{\ell^-}^n$ & $C_{nr} + C_{rn}$ & P-Odd, CP-Even\\
        $\cos\theta_{\ell^+}^n\cos\theta_{\ell^-}^r-\cos\theta_{\ell^+}^r\cos\theta_{\ell^-}^n$ & $C_{nr} - C_{rn}$ & P-Odd, CP-Odd\\
        $\cos\theta_{\ell^+}^n\cos\theta_{\ell^-}^k+\cos\theta_{\ell^+}^k\cos\theta_{\ell^-}^n$ & $C_{nk} + C_{kn}$ & P-Odd, CP-Even\\
        $\cos\theta_{\ell^+}^n\cos\theta_{\ell^-}^k-\cos\theta_{\ell^+}^k\cos\theta_{\ell^-}^n$ & $C_{nk} - C_{kn}$ & P-Odd, CP-Odd\\
        \hline
    \end{tabular*}
\end{table}

One can perform a fit against these 1-D function to finally obtain the relevant parameters of density matrix. An alternative method exists to obtain the parameters i) expectation value of of relevant angular function as
\begin{align}
    \langle f \rangle = \int d\cos\theta_{\ell^+}d\cos\theta_{\ell^-}f(\theta_{\ell^+},\theta_\ell^-)\Big(\frac{1}{\sigma}\frac{d^2\sigma}{d\cos\theta_{\ell^+}d\cos\theta_{\ell^-}}\Big),
\end{align}
which implies,
\begin{align}
    B_i^\pm = \frac{3}{\alpha_{\ell^\pm}} \langle f(\theta_{\ell^+},\theta_{\ell^-}) \rangle, \qquad C_{ij} = \frac{9}{\alpha_{\ell^+}\alpha_{\ell^-}} \langle f(\theta_{\ell^+},\theta_{\ell^-})\rangle.
\end{align}
and ii) asymmetries of angular functions~\cite{Boudjema:2009fz,Rahaman:2023pte}
\begin{align}
B_i^{\pm} 
&= \frac{2}{\alpha_{\ell^\pm}}
\frac{
N\!\left(f(\theta_{\ell^+},\theta_{\ell^-}) > 0\right)
-
N\!\left(f(\theta_{\ell^+},\theta_{\ell^-}) < 0\right)
}{
N\!\left(f(\theta_{\ell^+},\theta_{\ell^-}) > 0\right)
+
N\!\left(f(\theta_{\ell^+},\theta_{\ell^-}) < 0\right)
},
\nonumber\\[6pt]
C_{ij}
&= \frac{4}{\alpha_{\ell^+}\alpha_{\ell^-}}
\frac{
N\!\left(f(\theta_{\ell^+},\theta_{\ell^-}) > 0\right)
-
N\!\left(f(\theta_{\ell^+},\theta_{\ell^-}) < 0\right)
}{
N\!\left(f(\theta_{\ell^+},\theta_{\ell^-}) > 0\right)
+
N\!\left(f(\theta_{\ell^+},\theta_{\ell^-}) < 0\right)
}.
\end{align}
The relevant angular functions $f(\theta_{\ell^+},\theta_{\ell^-})$ and its associated parameters of density matrix along with its discrete symmetry characterization are shown in Table~\ref{tab:obs} following Ref.~\cite{CMS:2019nrx}. In the current work, we stick to method of asymmetries, while others are listed for completeness. Once these quantities are reconstructed using the one of the above listed method, the full density matrix can be assembled and subsequently employed to evaluate the quantum-correlation measures discussed in the next section.

\section{Non-classicality in top anti-top quark system}
\label{sec:corr}

The spin state of the $t\bar t$ system constitutes an effective two-qubit mixed state described in terms of  the density matrix $\rho_{t\bar t}$ introduced in Sec.~\ref{sec:formalism}. Non-classical correlations in this bipartite system can be quantified through QE  measures, QDs, and related quantum information quantities all of which can be expressed directly in terms of the Bloch vectors $A_i$, $B_i$ and correlation matrix $C_{ij}$.

\noindent
\textbf{Concurrence:} For a general two-qubit mixed state, QE can be quantified using the concurrence~\cite{Hill:1997pfa},
\begin{equation}
\mathcal{C} = \max\left(0,\lambda_1-\lambda_2-\lambda_3-\lambda_4\right),
\end{equation}
where $\lambda_i$'s are the square roots of the eigenvalues, in decreasing order, of the matrix
\begin{equation}
R = \rho_{t\bar t} \,(\sigma_y \otimes \sigma_y)\,\rho_{t\bar t}^*\,(\sigma_y \otimes \sigma_y).
\end{equation}
The concurrence satisfies $0 \le \mathcal{C} \le 1$, with $\mathcal{C}=0$ corresponding to separable states and $\mathcal{C}=1$ to maximally entangled states.
\\
\textbf{Quantum discord:} To quantify non-classical correlations beyond QE, we employ QD~\cite{Luo:2008ecu}. For a bipartite state $\rho_{AB}$, the total correlations are measured as the quantum mutual information
\begin{equation}
\mathcal{I}(\rho_{AB}) = S(\rho_A) + S(\rho_B) - S(\rho_{AB}),
\end{equation}
where $S(\rho) = -\mathrm{Tr}(\rho \log_2 \rho)$ is the von Neumann entropy. Classical correlations are defined as
\begin{equation}
\mathcal{J}_A(\rho_{AB}) = S(\rho_B) - \min_{\{\Pi_k^A\}} 
\sum_k p_k\, S(\rho_{B|k}),
\end{equation}
where the minimization runs over all projective measurements $\{\Pi_k^A\}$ on subsystem $A$. The quantum discord (with measurement on $A$) is then
\begin{align}
\label{eq:discord}
\mathcal{D}_A(\rho_{AB}) &= \mathcal{I}(\rho_{AB}) - \mathcal{J}_A(\rho_{AB}), \nonumber\\
&= S(\rho_B) - S(\rho_{AB}) + \min_{\{\Pi_k^A\}} 
\sum_k p_k\, S(\rho_{B|k}).
\end{align} 
The QD is in general asymmetric i.e, $\mathcal{D}_A(\rho_{AB}) \neq \mathcal{D}_B(\rho_{AB})$. This inequality can be a test for CP~\cite{Afik:2022dgh}. Within the SM, $t\bar{t}$ production at LHC respects CP at the leading order $B^+_i = B_i^-$, thus, QD remains symmetric.

The evaluation of QD as defined in Eq.~\eqref{eq:discord} generally requires a nontrivial numerical minimization over all possible measurement bases, making it computationally demanding and often  analytically intractable. The closed-form analytical expressions are known only for certain restricted classes of states. A particularly important class is  states with maximally mixed marginals, for which the single-party reduced density matrices are the identity. This condition is equivalent to the vanishing of both Bloch vectors and allows the state to be fully characterized in terms  the spin correlation matrix alone. For such states,  derived a closed-form analytical expression for QD was derived by showing that the optimization over measurement bases reduces to identifying the largest singular value of correlation matrix~\cite{Luo:2008ecu}. Thus, in the SM at the leading order, the $t\bar{t}$ becomes a state with maximally mixed marginals.

Nevertheless, the presence of anomalous physics could break the CP invariance and QD calculation requires the non-trivial minimization. To circumvent the minimization problem, we work with the geometric   quantum discord  (GQD) defined as~\cite{Dakic:2010xfz}
\begin{equation}
\label{eq:gqd1}
\mathcal{D}^g_{A}(\rho) = \min_{\chi \in \Omega_0} \, \lVert \rho - \chi \rVert^2,
\end{equation}
where $\Omega_0$ denotes the set of zero-discord states and $\lVert X - Y \rVert^2 = \mathrm{Tr}\!\left[(X - Y)^2\right]$ is the squared Hilbert--Schmidt norm\footnote{The GQD expressed in Eq.~\eqref{eq:gqd1} as Hilbert--Schmidt norm is known to be non-contractive under certain quantum operations~\cite{Piani:2012ajz}}.  The closed form expression for $\mathcal{D}^g_A(\rho)$ in the case of two qubit system is given as~\cite{Dakic:2010xfz}
\begin{align}
    \mathcal{D}^g_A(\rho) = \frac{1}{4}\Big[\lVert \Vec{x}\rVert^2 + \lVert C\rVert^2 - \kappa_{\mathrm{max}}\Big],
\end{align}
where $\Vec{x}$ is the Bloch vector associated with the first party, $C$ is the correlation matrix, $\lVert C\rVert^2 = \mathrm{Tr}\Big[C^TC\Big]$ and $\kappa_{\mathrm{max}}$ is the largest eigenvalue of matrix $K = \Vec{x}\Vec{x}^T + CC^T$.
\\
\noindent
\textbf{Bell nonlocality:}  For local hidden-variable theories, the Clauser--Horne--Shimony--Holt (CHSH) inequality can be written as~\cite{Clauser:1969ny}
\begin{equation}
\label{eq:CHSH}
\mathcal{B}(\rho) \equiv \left|
\vec a_1 \cdot C \cdot (\vec b_1 - \vec b_2)
+
\vec a_2 \cdot C \cdot (\vec b_1 + \vec b_2)
\right|
\le 2,
\end{equation}
where $\vec a_1$ and $\vec a_2$ denote two normalized spatial directions along which the top-quark spin is measured, while $\vec b_1$ and $\vec b_2$ correspond to the measurement directions for the anti-top quark. The matrix $C$ represents the spin-correlation matrix of the $t\bar{t}$ system. Finding the largest violation of Bell's
inequality corresponds to choosing the set of spatial directions that maximizes the Bell variable $\mathcal{B}(\rho)$. The optimization procedure yields the maximal violation of the CHSH inequality for a given two-qubit state $\rho$ as
\begin{equation}
\mathcal{B}(\rho)
=
\max_{\vec a_1, \vec a_2, \vec b_1, \vec b_2}
\left|
\vec a_1 \cdot C \cdot (\vec b_1 - \vec b_2)
+
\vec a_2 \cdot C \cdot (\vec b_1 + \vec b_2)
\right|.
\end{equation}
This maximum can be evaluated analytically as~\cite{Fabbrichesi:2021npl,Horodecki:1995nsk}
\begin{equation}
\mathcal{B}(\rho)
=
2 \sqrt{\mu_1 + \mu_2},
\end{equation}
where $\mu_1$ and $\mu_2$ are the two largest eigenvalues of the matrix $C^{\mathrm T} C$, with $C$ denoting the spin-correlation matrix of the $t\bar{t}$ system.

\section{SMEFT description of top-quark pair production at the LHC}
\label{sec:smeft}

Top-quark pairs are produced at the LHC primarily via QCD interactions, with subleading contributions from electroweak processes. Deviations from the Standard Model (SM) arising from heavy new physics at a scale $\Lambda \gg v$ can be systematically described within the framework of SMEFT,  where the SM Lagrangian is extended by higher-dimensional operators constructed from SM fields and invariant under the gauge group $SU(3)_C \times SU(2)_L \times U(1)_Y$ as~\cite{Buchmuller:1985jz}
\begin{equation}
\mathcal{L}_{\mathrm{SMEFT}} = \mathcal{L}_{\mathrm{SM}} 
+ \sum_i \frac{C_i^{(5)}}{\Lambda} \mathcal{O}_i^{(5)} 
+ \sum_i \frac{C_i^{(6)}}{\Lambda^2} \mathcal{O}_i^{(6)} 
+ \mathcal{O}\Big(\frac{1}{\Lambda^3}\Big),
\end{equation}
where $C_i^{(d)}$ denote the Wilson coefficients (WCs) of the dimension-$d$ operators. This expansion in inverse powers of $\Lambda$ formally includes all operators consistent with the SM gauge symmetries. The unique dimension-5 Weinberg operator violates baryon number conservation~\cite{Weinberg:1979sa}. Hence, dimension-6  operators provide the leading-order contributions relevant for LHC observables.

The $t\bar{t}$ process provides a sensitive probe of higher-dimensional operators within the SMEFT framework. In the present work, we focus on the subset of operators that induce anomalous contributions to the SM $gt\bar{t}$ and $t\bar{t}Z$ vertices, which primarily affect the production dynamics of top-quark pairs. While SMEFT effects can also appear in top decays, these contributions are strongly constrained~\cite{Fabbrichesi:2014wva,Cao:2015doa,Bernreuther:2015yna} and thus have a minimal impact on measured distributions~\cite{Bernreuther:2015yna,Baumgart:2012ay}. The independent dimension-six operators relevant for the $gt\bar{t}$ and $t\bar{t}Z$ vertices in the Warsaw basis are~\cite{Grzadkowski:2010es}
\begin{align}
\mathcal{O}_{uW} &= (\bar q_p \sigma^{\mu\nu} u_r)\,\tau^I \widetilde{\Phi}\, W_{\mu\nu}^I, &
\mathcal{O}_{uB} &= (\bar q_p \sigma^{\mu\nu} u_r)\, \widetilde{\Phi}\, B_{\mu\nu}, \nonumber\\
\mathcal{O}_{\Phi u} &= (\Phi^\dagger i \overleftrightarrow{D}_\mu \Phi)(\bar u_p \gamma^\mu u_r), &
\mathcal{O}_{\Phi q}^{(1)} &= (\Phi^\dagger i \overleftrightarrow{D}_\mu \Phi)(\bar q_p \gamma^\mu q_r), \nonumber\\
\mathcal{O}_{\Phi q}^{(3)} &= (\Phi^\dagger i \overleftrightarrow{D}_\mu^I \Phi)(\bar q_p \tau^I \gamma^\mu q_r), &
\mathcal{O}_{uG} &= (\bar q_p \sigma^{\mu\nu} T^a u_r)\, \widetilde{\Phi}\, G_{\mu\nu}^a,
\label{eq:dim6_ops}
\end{align}
where $q_p$ and $u_r$ denote the left-handed quark doublets and right-handed up-type quark singlets, respectively; $\Phi$ is the Higgs doublet with $\widetilde{\Phi} = i\tau^2 \Phi^*$. The $SU(2)_L$ and $SU(3)_C$ generators are $\tau^I$ and $T^a = \lambda^a/2$ where $\lambda^a$ are the Gell--Mann matrices and the derivative structures are
\begin{align}
\Phi^\dagger i \overleftrightarrow{D}_\mu^I \Phi &= i \Phi^\dagger \tau^I D_\mu \Phi - i (D_\mu \Phi)^\dagger \tau^I \Phi,\\
D_\mu &= \partial_\mu + i \frac{g_W}{2} \tau^I W_\mu^I + i \frac{g_1}{2} B_\mu,
\end{align}
with the gauge field-strength tensors
\begin{align}
W_{\mu\nu}^I &= \partial_\mu W_\nu^I - \partial_\nu W_\mu^I + g_W \epsilon^{IJK} W_\mu^J W_\nu^K, \nonumber\\
G_{\mu\nu}^a &= \partial_\mu G_\nu^a - \partial_\nu G_\mu^a + g_s f^{abc} G_\mu^b G_\nu^c.
\end{align}

In addition to these operators, top quark pair production is sensitive to modifications of the triple-gluon vertex, encoded by the CP-even dimension-6 operator
\begin{align}
\mathcal{O}_G = g_s f^{abc} G_\mu^{a\nu} G_\nu^{b\rho} G_\rho^{c\mu},
\end{align}
which has been studied extensively~\cite{Krauss:2016ely,Hirschi:2018etq,Goldouzian:2020wdq,Bardhan:2020vcl}. The first experimental limit on this operator, $\Lambda/\sqrt{C_G} > 5.2$~TeV, was derived from multijet events at the LHC using the transverse observable, $
S_T = \sum_{j=1}^{N_\mathrm{jets}} E_{T,j} + \slashed{E}_T\; (> 50~\mathrm{GeV}).
$
The CP-odd dual of $\mathcal{O}_G$ induces anomalous $ggg$ interactions but is strongly constrained by low-energy measurements of the neutron electric dipole moment~\cite{Dekens:2013zca}.  Finally, top quark pair production is also impacted by four-fermion operators, which contribute via $q\bar q \to t\bar t$ channels~\cite{Buckley:2015nca,Bernreuther:2015yna,Zhang:2010dr}. Their effects on spin correlations and Bloch vectors have been computed in details~\cite{Bernreuther:2015yna}, providing additional handles to probe SMEFT contributions in the top sector.

The most general Lagrangian with both CP-even and CP-odd physics in the $t\bar{t}Z$ and $gt\bar{t}$ vertices  can be written as~\cite{Aguilar-Saavedra:2008nuh,Ravina:2021kpr,Aguilar-Saavedra:2012ehg}
\begin{align}
\label{eq:anom_lag}
\mathcal{L}_{\mathrm{eff}} &= e\,\bar{t}\Big[\gamma^\mu \left(\Delta C_1^V + \gamma_5 \Delta C_1^A\right)+ \frac{i\sigma^{\mu\nu} q_\nu}{m_Z}\left( C_2^V + i\gamma_5  C_2^A\right)
\Big] t\, Z_\mu \nonumber\\&\quad - g_s \bar{t} \gamma^\mu G_\mu^a \frac{\lambda^a}{2} t- \frac{g_s}{2 m_t}\,
\bar{t}\sigma^{\mu\nu}\left(\hat{\mu}_t + i \hat{d}_t \gamma_5\right)
G_{\mu\nu}^a \frac{\lambda^a}{2} t .
\end{align}
Here, $e$ and $g_s$ denote the electromagnetic and strong coupling constants, respectively; $G_\mu^a$ is the gluon field, $G_{\mu\nu}^a$ its field-strength tensor. The momentum transfer carried by the $Z$ boson is denoted by $q_\nu$.

The coefficients $\Delta C_1^{V,A} \equiv C_1^{V,A} - C_1^{V_A}(\mathrm{SM})$ parametrize deviation in the vector and axial-vector $t\bar{t}Z$ interactions from the SM value. At tree level in the SM, the $C_1^{V,A}$ is a function of weak mixing angle ($\theta_W$), charge ($Q_f$) and third iso-spin of the fermion and is given by~\cite{ParticleDataGroup:2022pth}
\begin{equation}
C_{1,\mathrm{SM}}^V = \frac{I_{3,q}^f - 2Q_f \sin^2\theta_W}{2\sin\theta_W\cos\theta_W}\simeq 0.24, 
\qquad 
C_{1,\mathrm{SM}}^A  = \frac{-I_{3,q}^f}{2\sin\theta_W\cos\theta_W}\simeq -0.60,
\end{equation}
while $C_2^{V,A}=0$. The dipole couplings $C_2^V$ (CP-even) and $C_2^A$ (CP-odd) correspond to the weak magnetic and weak electric dipole moments of the top quark~\cite{Rontsch:2014cca,Schulze:2016qas,Aguilar-Saavedra:2008nuh}, respectively.

The parameters $\hat{\mu}_t$ and $\hat{d}_t$ denote the chromomagnetic dipole moment (CMDM) and chromoelectric dipole moment (CEDM) of the top quark. In the SM, $\hat{\mu}_t$ is generated at one loop from both quantum chromodynamics and electroweak~\cite{Aranda:2020tox,Choudhury:2014lna,Bermudez:2017bpx,Martinez:2007qf}, whereas $\hat{d}_t$ first arises at three-loop order via the complex phase of the CKM matrix and is therefore highly suppressed~\cite{Czarnecki:1997bu,Khriplovich:1985jr}. Any sizable deviation would signal physics beyond the SM.

After electroweak symmetry breaking, $\Phi \to (0,, (v+h)/\sqrt{2})^T$, the operators in Eq.~(\ref{eq:dim6_ops}) induce corrections to the effective $t\bar{t}Z$ couplings presented in Eq.~(\ref{eq:anom_lag}). These modifications take the form~\cite{Machethe:2025oev}
\begin{align}
\Delta C_1^V &= \frac{v^2}{\Lambda^2}\,\mathfrak{Re}\Big[-C_{\Phi u} - C_{\Phi u}^-\Big],
\qquad
\Delta C_1^A = \frac{v^2}{\Lambda^2}\,\mathfrak{Re}\Big[-C_{\Phi u} + C_{\Phi q}^-\Big],
\nonumber\\
C_2^V &= \frac{\sqrt{2}v^2}{\Lambda^2}C_{tZ}, 
\qquad
C_2^A = \frac{\sqrt{2}v^2}{\Lambda^2}C_{tZ}^I,
\end{align}
where $v= 246~\mathrm{GeV}$ denotes the Higgs field vacuum expectation value. The effective parameters appearing above are related to the WCs of Eq.~(\ref{eq:dim6_ops}) through~\cite{CMS:2019too,Aguilar-Saavedra:2018ksv}
\begin{align}
\label{eq:relation}
C_{\Phi q}^- &= C_{\Phi q}^{(1)} - C_{\Phi q}^{(3)},\nonumber \\
C_{tZ} &= \mathfrak{Re}\Big[-\sin\theta_W\,C_{uB} + \cos\theta_W\,C_{uW}\Big],
\nonumber\\
C_{tZ}^I &= \mathfrak{Im}\Big[-\sin\theta_W\,C_{uB} + \cos\theta_W\,C_{uW}\Big].
\end{align}
The real and imaginary parts of the WCs combination involving $C_{uB}$ and $C_{uW}$ generate, respectively, the weak magnetic and weak electric dipole moments of the top quark. Meanwhile, the coefficients $C_{\Phi u}$ and $C_{\Phi q}^-$ induce shifts in the neutral-current couplings of the top quark to the $Z$ boson.

The most stringent experimental constraints on these WCs are provided 
by the ATLAS Collaboration~\cite{ATLAS:2023eld} using $t\bar{t}Z$ production at $\sqrt{s} = 13~\mathrm{TeV}$ with $\mathcal{L}=140$ fb$^{-1}$. The independent one-parameter 95\% CL bounds on the relevant 
Warsaw-basis coefficients are found to be $C^{(1)}_{\Phi q} \in [-1.4,\, 0.84]$, 
$C^{(3)}_{\Phi q} \in [-0.95,\, +2.0]$, $\mathfrak{Re}[C_{uW}] \in 
[-1.0,\, +1.0]$, $\mathfrak{Im}[C_{uW}] \in [-0.84,\, +1.0]$, $\mathfrak{Re}[C_{uB}] = [-1.7,\,+1.6]$ and $\mathfrak{Im}[C_{uB}] = [-1.9,\,+1.9]$ all in 
units of $\mathrm{TeV}^{-2}$. Bounds on the derived combinations $\Delta C^{A,V}_1$ 
and $C^{A,V}_2$ entering Eq.~(\ref{eq:anom_lag}) can be obtained via the relations 
in Eq.~\eqref{eq:relation}. Additional studies investigating these anomalous $t\bar{t}Z$ couplings through different processes, observables, and collider scenarios can be found in Refs.~\cite{Rahaman:2022dwp,Machethe:2025oev,Buckley:2015nca,Durieux:2018tev,Cao:2020npb,Cao:2015qta} and references therein.

Following electroweak symmetry breaking, the chromomagnetic and chromoelectric dipole moments are related to $C_{uG}$ through~\cite{Aguilar-Saavedra:2008nuh}
\begin{align}
\hat{\mu}_t = \frac{v}{\sqrt{2}\,\Lambda^2}\,\mathfrak{Re}[C_{uG}], \qquad
\hat{d}_t = \frac{v}{\sqrt{2}\,\Lambda^2}\,\mathfrak{Im}[C_{uG}].
\end{align}
The CMS Collaboration~\cite{CMS:2019kzp} reports constraints on the top quark anomalous chromomagnetic ($\hat{\mu}_t$) and chromoelectric ($\hat{d}_t$) moments using proton-proton collisions at $\sqrt{s}=13$ TeV corresponding to an integrated luminosity of $35.9$ fb$^{-1}$. The forward-backward asymmetry is employed to constrain the anomalous couplings in $t\bar{t}$ events decaying to leptons+jets final states. The measured chromomagnetic moment is $\hat{\mu}_t = -0.024^{+0.013}_{-0.009}\,(\mathrm{stat})^{+0.016}_{-0.011},(\mathrm{syst})$, while a $95\%$ CL upper limit of $|\hat{d}_t| < 0.03$ is placed on the chromoelectric moment. Additional probes of anomalous $\hat{\mu}_t$ and $\hat{d}_t$ couplings can be found in the Refs.~\cite{Hioki:2009hm,Rahaman:2023pte}, and in references therein.

Next, we move on to discuss the collider analysis and the impact of anomalous couplings on QI quantities.

\section{Results}
\label{sec:results}
We restrict ourselves to a parton-level analysis of 
$t\bar t$ production and decay, providing a controlled environment in which 
the spin density matrix can be reconstructed directly from truth-level 
kinematics. This setup corresponds to the maximal information scenario and 
serves as a clean benchmark for assessing the intrinsic sensitivity of 
non-classical observables to anomalous top chromo dipole interactions. 
A fully realistic detector-level analysis including backgrounds, event 
reconstruction, and experimental systematics will be presented in a 
separate dedicated study.

Monte Carlo event generation is performed using 
\textsc{MadGraph5\_aMC@NLO} for the hard scattering process. We generate 
$pp \to t\bar t$ events at $\sqrt{s}=13$ TeV at leading order, followed by 
fully leptonic decays,
\[
t \to b \ell^+ \nu_\ell, 
\qquad 
\bar t \to \bar b \ell^- \bar\nu_\ell,
\]
with spin correlations between production and decay preserved at the 
amplitude level. Since the present study is restricted to parton level, 
we do not include parton showering, hadronization, or detector simulation. 
All QI quantities considered are constructed directly from truth-level four-momenta. 

We analyze the spin-density matrix in both the helicity ($k$-$r$-$n$) basis, and study its behavior across different invariant-mass regions 
of the $t\bar t$ system:
\begin{enumerate}
    \item \textbf{Threshold region (M1)}: $m_{t\bar{t}} \le 400$~GeV,
    \item \textbf{Intermediate region (M2)}: $400 < m_{t\bar{t}} \le 800$~GeV,
    \item \textbf{Mildly boosted region (M3)}: $800 < m_{t\bar{t}} \le 1000$~GeV,
    \item \textbf{Extreme boosted region (M4)}: $m_{t\bar{t}} > 1000$~GeV.
\end{enumerate}
These regions are chosen to systematically probe the dependence of spin correlations and quantum mixedness on the kinematic boost of the system, following the binning convention of $m_{t\bar{t}}$ as in Ref.~\cite{Han:2024ugl}. We discuss our results in the next section.

\begin{table}[!t]
    \centering
    \caption{Details of the nine spin correlation matrix elements along with the concurrence measure in the $k$-$r$-$n$ helicity basis at parton level, for different $t\bar{t}$ invariant-mass regions within the Standard Model. Event counts for each region are also reported.}
    \label{tab:parton_krn}
    \scriptsize
    \renewcommand{\arraystretch}{1.4}
    \begin{tabular}{lccccccccccc}
    \hline
      Region & Events & $C_{kk}$ & $C_{kr}$ &$C_{kn}$ & $C_{rk}$ &$C_{rr}$ & $C_{rn}$ & $C_{nk}$ & $C_{nr}$ &$C_{nn}$ & $\mathcal{C}[\rho_{t\bar{t}}]$    \\
      \hline
       Un-binned & $20\times 10^6$ & $-0.339$ & $0.112$ & $0.0$ & $0.113$ & $-0.014$ & $0.0$ & $0.0$ & $0.0$ & $-0.334$ & $0.0$ \\
       $m_{t\bar{t}} \le 400$~GeV & $4.24\times 10^6$ & $-0.590$ & $0.096$ & $0.0$ & $0.094$ & $-0.314$ & $0.0$ & $0.0$ & $0.0$ & $-0.538$ & $0.221$\\
       $400 < m_{t\bar{t}} \le 800$~GeV & $14.39\times 10^6$ & $-0.307$ & $0.119$ & $0.0$ & $0.122$ & $0.056$ & $0.0$ & $0.0$ & $0.0$ & $-0.287$ &$0.0$\\
       $800 < m_{t\bar{t}} \le 1000$~GeV & $8.64\times 10^5$ & $0.059$ & $0.101$ & $0.0$ & $0.098$ & $0.185$ & $0.0$ & $0.0$ & $0.0$ & $-0.202$& $0.0$\\
       $m_{t\bar{t}} > 1000$~GeV & $5.0\times 10^5$ & $0.195$ & $0.074$ & $0.0$ & $0.075$ & $0.170$ & $0.0$ & $0.0$ & $0.0$ & $-0.169$& $0.0$ \\
       \hline
    \end{tabular}
\end{table}

\subsection{Correlations in Standard Model: Helicity basis} 
Table~\ref{tab:parton_krn} summarizes the nine independent elements of the spin correlation matrix in the $k$-$r$-$n$ helicity basis at parton level within  SM, together with the number of events in each invariant-mass region and the corresponding concurrence $\mathcal{C}[\rho_{t\bar{t}}]$ quantifying QE of the $t\bar{t}$ system.

In the $k$-$r$-$n$ helicity basis (Table~\ref{tab:parton_krn}), the correlation matrix is predominantly diagonal, with $C_{kr}$ and $C_{rk}$ showing small but non-zero values, consistent with CP invariance in $t\bar{t}$ production. All other off-diagonal elements are negligible across the full invariant-mass range. The diagonal entries are largest in the threshold region ($m_{t\bar{t}} \le 400$~\text{GeV}), indicating strong spin alignment for slowly moving top-quark pairs. As $m_{t\bar{t}}$ increases, the diagonal components decrease in magnitude and may change sign, reflecting the impact of Lorentz boosts on the spin configuration and the increasing mixedness of the two-qubit state. A non-vanishing concurrence is observed only in the threshold region, where the strong diagonal correlations lead to a genuinely entangled spin state. In all higher invariant-mass bins the concurrence vanishes within numerical precision, indicating that the corresponding density matrices are separable despite retaining non-zero classical spin correlations. 

In the helicity basis, with approximate CP invariance of the SM, the correlation matrix reduces to 
\begin{align}
    C_{ij} = \begin{pmatrix}
        C_{kk} & C_{kr} &0 \\
        C_{kr} & C_{rr} & 0\\
        0 &0 &C_{nn}
    \end{pmatrix}.
\end{align}
then, the form of the quantum discord for $t\bar{t}$ system in the helicity basis reduces to~\cite{Han:2024ugl}
\begin{align}
    \mathcal{D}_t(t\bar{t}) &= 1 + \frac{1}{4}\Big(1-\sum_i C_{ii}\Big)\,\mathrm{log}_2\Big(\frac{1-\sum_iC_{ii}}{4}\Big) \nonumber\\
    &+ \frac{1}{4}\Big(1+C_{kk}-C_{nn}+C_{rr}\Big)\mathrm{log}_2\Big(\frac{1+C_{kk}-C_{nn}+C_{rr}}{4}\Big)\nonumber\\
    &+ \frac{1}{4}\Big(1+C_{nn}-\Delta\Big)\,\mathrm{log}_2\Big(\frac{1+C_{nn}-\Delta}{4}\Big)
    + \frac{1}{4}\Big(1+C_{nn}+\Delta\Big)\,\mathrm{log}_2\Big(\frac{1+C_{nn}+\Delta}{4}\Big)\nonumber\\
    &- \frac{1}{2}(1+\lambda)\,\mathrm{log}_2\Big(\frac{1+\lambda}{2}\Big)
    -\frac{1}{2}(1-\lambda)\,\mathrm{log}_2\Big(\frac{1-\lambda}{2}\Big),
\end{align}
where $\Delta = \sqrt{C_{kk}^2+4\,C_{kr}^2+C_{rr}^2-2\,C_{kk}C_{rr}}$ and $\lambda = \mathrm{max}\Big(|C_{nn}|,\frac{1}{2}|C_{kk}+C_{rr}-\Delta|, \frac{1}{2}|C_{kk}+C_{rr}+\Delta|\Big).$
\begin{figure}[!htb]
    \centering
    \includegraphics[width=0.49\textwidth]{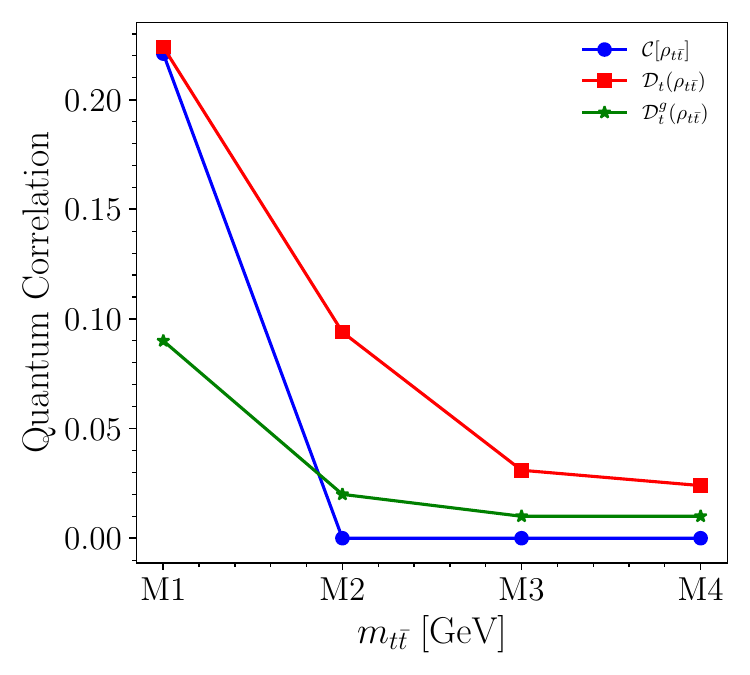}
    \includegraphics[width=0.49\textwidth]{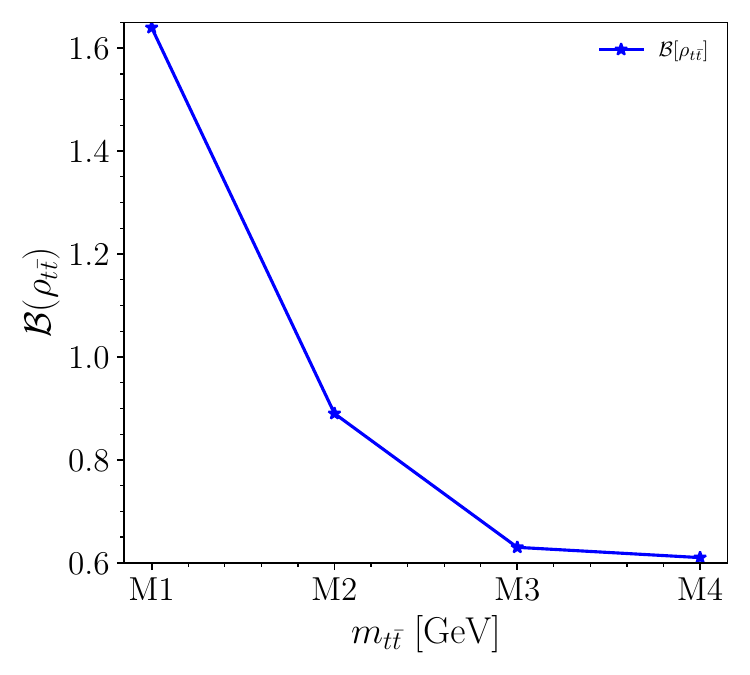}
    \caption{Distribution for concurrence measure of quantum entanglement and quantum discord in the left panel and Bell variable in the right panel for $t\bar{t}$ process at $13$ TeV LHC. The distribution are obtained at the parton level for the di-leptonic decay of top quarks in the $k$-$r$-$n$ helicity basis.}
    \label{fig:parton_quant}
\end{figure}

In the left panel of Fig.~\ref{fig:parton_quant}, we display the invariant-mass dependence of the 
concurrence $\mathcal{C}[\rho_{t\bar{t}}]$, QD 
$\mathcal{D}_t(\rho_{t\bar{t}})$, and GQD 
$\mathcal{D}_t^g(\rho_{t\bar{t}})$ for the $t\bar{t}$ system at parton level. 
In the threshold region (M1), the concurrence and quantum discord coincide, 
indicating that the reconstructed density matrix approaches a pure state in the 
limit $\beta \to 0$. This behaviour arises from the dominance of the 
$gg$-initiated production channel near threshold. As seen from 
Eqs.~\eqref{eq:qq} and~\eqref{eq:gg}, the $q\bar{q}$ channel yields a mixed 
spin configuration, whereas the $gg$ channel produces a pure $t\bar{t}$ spin 
state in this limit. Consequently, the concurrence attains its maximal value in 
the threshold region. As $m_{t\bar{t}}$ increases (M2--M4), the kinematic boosts 
induce additional mixing in the spin degrees of freedom. The concurrence 
correspondingly decreases and vanishes in the intermediate and boosted regions, 
signaling that the density matrix becomes separable.

In contrast, both QD $\mathcal{D}_t$ and GQD  $\mathcal{D}_t^g$ remain positive across all invariant-mass bins. Since 
these measures capture non-classical correlations beyond QE, the 
conditions $\mathcal{D}_t(\rho_{t\bar{t}}) > 0$ and 
$\mathcal{D}_t^g(\rho_{t\bar{t}}) > 0$ demonstrate that even when the state is 
separable, the $t\bar{t}$ system retains genuinely quantum correlations. Notably, 
the GQD $\mathcal{D}_t^g$ exhibits qualitatively similar 
behaviour to the standard QD $\mathcal{D}_t$, remaining positive 
across all invariant-mass bins. Since the two measures are defined on different scales, the latter being entropic and the former a squared Hilbert-Schmidt distance, a direct quantitative comparison between them is not meaningful, however, both consistently signal the presence of non-classical correlations in 
the separable regime. The figure therefore illustrates the transition from an 
entanglement-dominated regime near threshold to separable yet non-classically 
correlated states at higher invariant mass, with both discord measures serving 
as sensitive probes of this residual quantum character.

In the right panel of Fig.~\ref{fig:parton_quant}, the maximal CHSH-Bell variable $\mathcal{B}$ remains below the classical bound $B = 2$ across all invariant-mass bins. While the $t\bar{t}$ system is entangled near threshold and retains non-classical correlations throughout, as indicated by the positive QD, these correlations are insufficient to violate the Bell inequality in the helciity basis. Our result is  similar to previous one~\cite{Han:2024ugl}. Next, we discuss the complementary result in beam basis.

\begin{table}[!t]
    \centering
    \caption{Spin correlation matrix elements in the beam basis at parton level for various $t\bar{t}$ invariant-mass regions within the Standard Model. Diagonal and off-diagonal entries are listed along with the number of events in each region.}
    \label{tab:parton_beam}
    \scriptsize
    \renewcommand{\arraystretch}{1.3}
    \begin{tabular}{lccccccccccc}
    \hline
    Region & Events & $C_{xx}$ & $C_{xy}$ &$C_{xz}$ & $C_{yx}$ &$C_{yy}$ & $C_{yz}$ & $C_{zx}$ & $C_{zy}$ &$C_{zz}$ & $\mathcal{C}[\rho_{t\bar{t}}]$    \\
    \hline
     Un-binned & $20\times 10^6$ & $-0.321$ &$0.0$& $0.0$& $0.0$ & $-0.322$ & $0.0$ & $0.0$ & $0.0$ & $-0.042$&$0.0$ \\
    $m_{t\bar{t}} \le 400$~GeV & $4.2\times 01^6$ &$-0.573$ &$0.0$ &$0.0$ &$0.0$ & $-0.574$&$0.0$ &$0.0$ &$0.0$ & $-0.293$ & $0.220$ \\
    $400 < m_{t\bar{t}} \le 800$~GeV &$14.4\times 10^6$ & $-0.273$ & $0.0$ & $0.0$ & $0.0$ & $-0.274$ & $0.0$&$0.0$ & $0.0$  & $0.011$ &$0.0$ \\
    $800 < m_{t\bar{t}} \le 1000$~GeV &$8.6\times 10^5$ & $-0.063$ & $0.0$ & $0.0$ & $0.0$ & $-0.061$ & $0.0$ & $0.0$ & $0.0$ & $0.172$ & $0.0$ \\
    $m_{t\bar{t}} > 1000$~GeV &$5.0\times 10^5$ & $-0.013$ &$0.0$ & $0.0$ & $0.0$ & $-0.009$ & $0.0$ & $0.0$ &$0.0$  &$0.218$ &$0.0$\\       
    \hline
    \end{tabular}
\end{table}
\subsection{Correlations in Standard Model: Beam basis}
In the beam-basis, the $z$-axis aligns with the proton beams, and the $x$- and $y$-axes are fixed according to the CMS coordinate system: the origin is at the nominal collision point, $x$ points radially inward toward the LHC center, and $y$ points vertically upward. 
Angles are defined with respect to these axes: $\phi$ from $x$ in the transverse plane, $\theta$ from $z$. 
Top quarks and their decay products are actively boosted first into the $t\bar{t}$ CM frame and then into the individual top/anti-top rest frames. 
Unlike the helicity basis, no event-dependent rotations are applied, and the spin correlation matrix is diagonal due to the beam-axis symmetry.

The beam basis (Table~\ref{tab:parton_beam}) shows a similar pattern to that of helicity one: diagonal elements $C_{xx} = C_{yy} = C_\perp$, and $C_{zz}$ dominate, with all off-diagonal components effectively suppressed. As in the helicity case, the threshold region exhibits the strongest spin alignment along the beam directions, and the corresponding entanglement measure $\mathcal{C}[\rho_{t\bar{t}}] \simeq 0.2$ is identical to that of the helicity basis, confirming entangled top pairs at threshold. With increasing $m_{t\bar{t}}$, the diagonal elements decrease in magnitude, and $C_{nn}$ changes sign in boosted bins, while $\mathcal{C}[\rho_{t\bar{t}}]$ approaches zero for both intermediate and boosted regions.

\begin{figure}[!htb]
    \centering
    \includegraphics[width=0.98\textwidth]{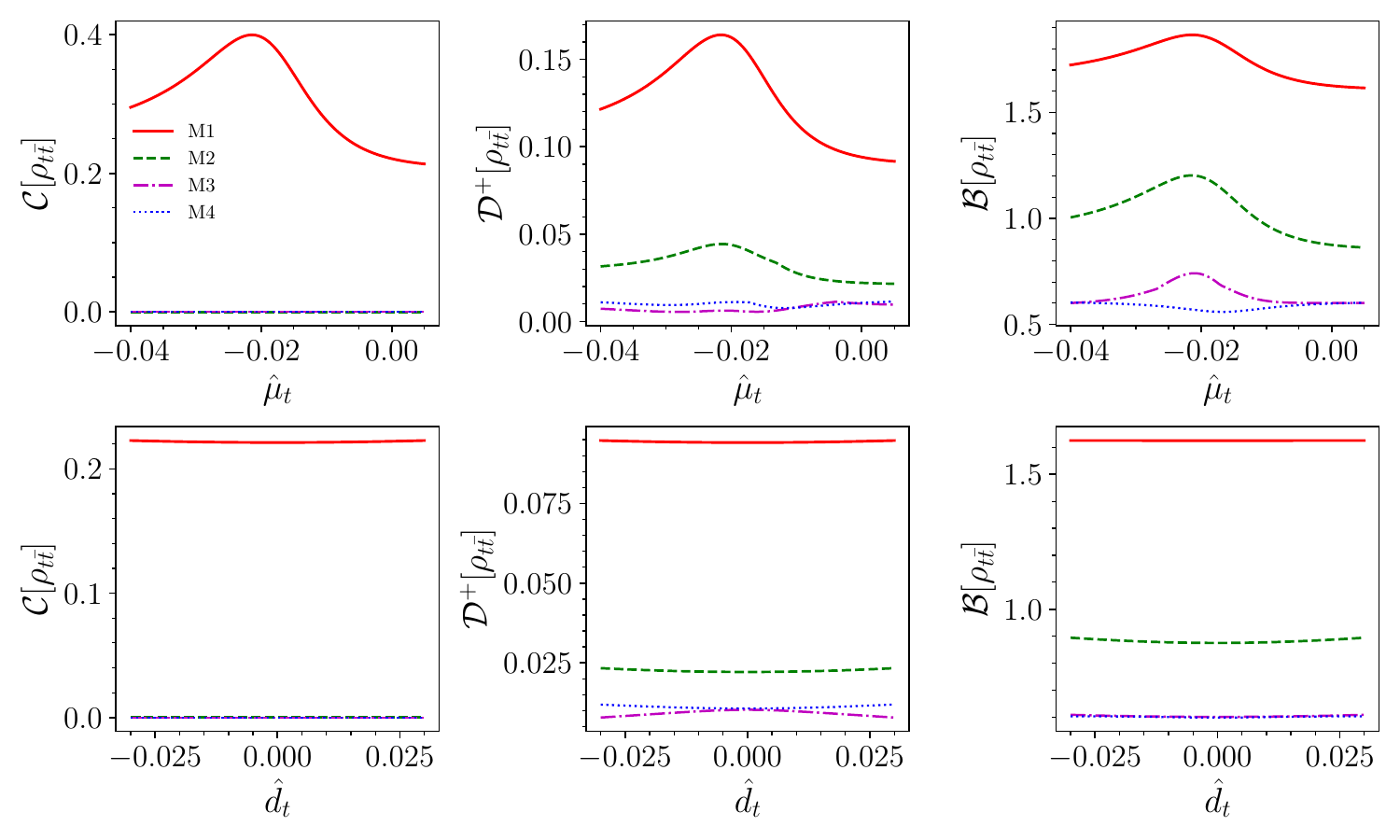}
    \caption{Distribution of concurrence, geometric quantum discord and Bell variable as a function of CP-even $\hat{\mu}_t$ and CP-odd $\hat{d}_t$ anomalous couplings. The distribution are shown at the parton level for $pp\to t\bar{t}$ followed by leptonic decay in four binned region of $m_{t\bar{t}}$ at $13$ TeV LHC.}
    \label{fig:oned_quant}
\end{figure}

\subsection{Non-classicality with anomalous chromo-dipole moments} 
Figure~\ref{fig:oned_quant} shows the dependence of the concurrence 
$\mathcal{C}[\rho_{t\bar{t}}]$, GQD  
$\mathcal{D}^+[\rho_{t\bar{t}}]$\footnote{Within the SM, CP invariance enforces 
$B^+_i = B^-_i = 0$, so that $\mathcal{D}^+ = \mathcal{D}^-$ identically, and the 
choice of subsystem on which the measurement is performed carries no deeper significance.  In the presence of anomalous CP-violating couplings, $\mathcal{D}^+$ and $\mathcal{D}^-$ 
become distinct; their difference is discussed in the later part.}, and Bell parameter 
$\mathcal{B}[\rho_{t\bar{t}}]$ on the CP-even coupling $\hat{\mu}_t$ (top row) and 
the CP-odd coupling $\hat{d}_t$ (bottom row), for four $m_{t\bar{t}}$ bins (M1--M4). 
Throughout this analysis, both couplings are varied strictly within their current 
experimentally allowed ranges: $\hat{\mu}_t \in [-0.04, +0.005]$ and 
$|\hat{d}_t| \leq 0.03$, consistent with the CMS 95\% CL constraints.  The distributions are shown at the parton level for $t\bar{t}$ production followed by its leptonic decay at $13$ TeV LHC.

On the top row, all three observables display a pronounced and asymmetric dependence 
on $\hat{\mu}_t$, driven by the linear interference of the chromo-magnetic dipole 
operator with the SM amplitude. In the threshold bin M1, 
$\mathcal{C}[\rho_{t\bar{t}}]$ develops a sharp peak of $\sim 0.4$ at 
$\hat{\mu}_t \approx -0.025$, before decreasing toward both the SM point and the 
lower edge of the allowed range. Notably, this peak coincides with the central value 
of the CMS measurement, implying that the current best-fit chromo-magnetic dipole 
moment would predict an enhancement of QE in the threshold region 
relative to the pure SM prediction. The GQD mirrors this behaviour, peaking at the 
same coupling value and falling off smoothly. The asymmetry of the distributions 
within the displayed range is a direct manifestation of the linear interference term: 
constructive interference with the SM amplitude occurs for negative $\hat{\mu}_t$, 
enhancing the quantum correlations relative to the SM prediction. The Bell parameter 
in M1 ranges between $1.6$ and $1.85$ across the allowed range, remaining below the 
classical bound of $2$ throughout, indicating that Bell inequality violation does not 
occur in the threshold bin for CP-even deformations within current experimental 
limits.

The higher bins M2--M4 show a systematic and sharp reduction in sensitivity. 
$\mathcal{C}[\rho_{t\bar{t}}]$ is essentially vanishing across the entire displayed 
range for M3 and M4, confirming that QE is strongly suppressed in 
the boosted regime. In contrast, $\mathcal{D}^+[\rho_{t\bar{t}}]$ remains finite 
for all bins even where $\mathcal{C} \simeq 0$, confirming the presence of 
non-classical correlations beyond QE. The Bell parameter in M2--M4 is 
nearly flat and significantly below $2$, clustered around $0.55$--$0.65$ for M3 
and M4, with no indication of Bell violation within the experimentally allowed 
region.

The bottom row displays the $\hat{d}_t$ dependence within the experimentally allowed 
region $|\hat{d}_t| \leq 0.03$. All distributions are strictly symmetric about 
$\hat{d}_t = 0$, as required by the CP-odd nature of the chromo-electric dipole 
operator, which enters CP-even observables only quadratically. Within this range, 
the three observables show a mild but visible response to $\hat{d}_t$. In all region of $m_{t\bar{t}}$, 
$\mathcal{C}[\rho_{t\bar{t}}]$ exhibits no variation. In M1, the GQD remains flat at the SM value, while for higher mass regions, it shows a mild change near $\hat{d}_t=0$ region.  The Bell parameter in M1 is nearly constant at $\sim 1.5$--$1.6$, showing minimal sensitivity to $\hat{d}_t$ in the threshold 
region. In the higher bins M2--M4, the Bell parameter show a 
slightly more pronounced but still modest variation across bins. Importantly, the 
Bell parameter remains well below the classical bound of $2$ across all bins and 
for all values of $\hat{d}_t$ within the experimentally allowed region, indicating 
that Bell inequality violation cannot be achieved for CP-odd chromo-electric dipole 
couplings consistent with current experimental constraints.

\begin{figure}[!htb]
    \centering
    \includegraphics[width=0.32\textwidth]{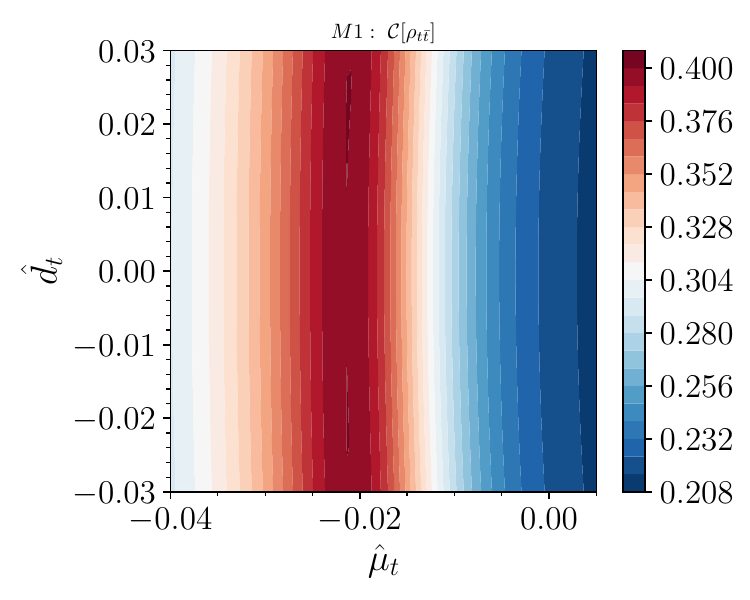}
    \includegraphics[width=0.32\textwidth]{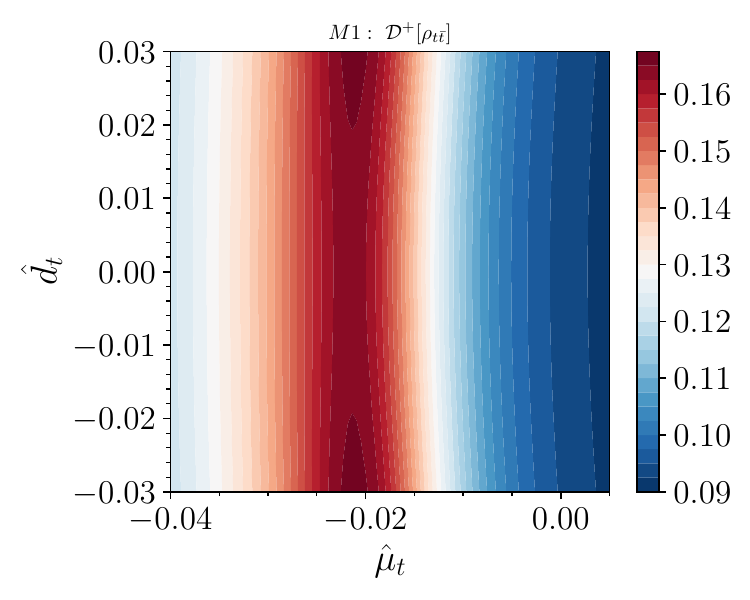}
    \includegraphics[width=0.32\textwidth]{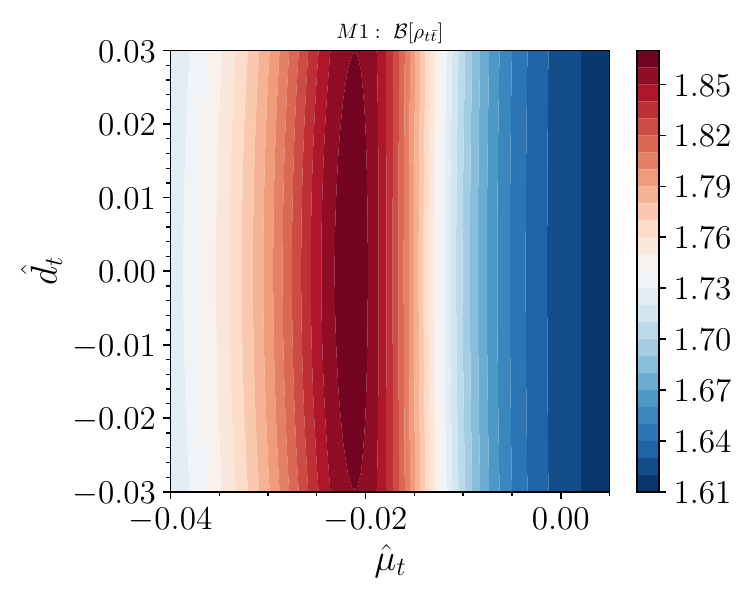}
    \includegraphics[width=0.32\textwidth]{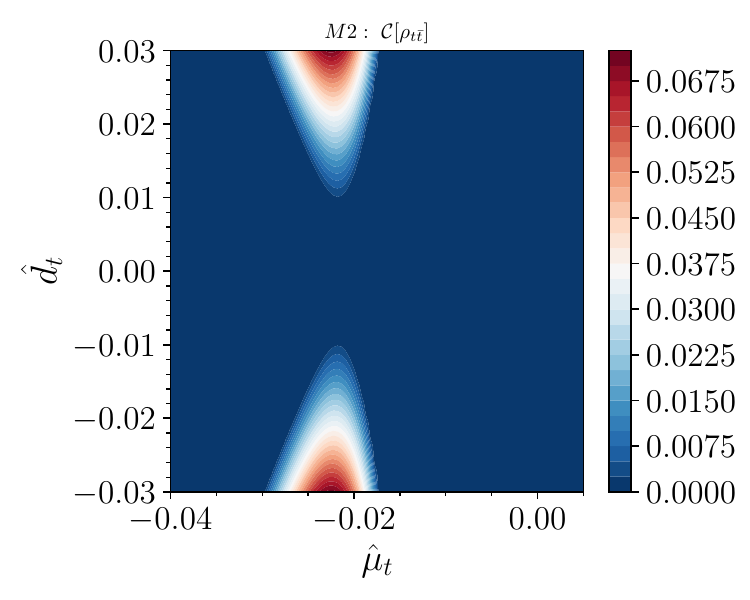}
    \includegraphics[width=0.32\textwidth]{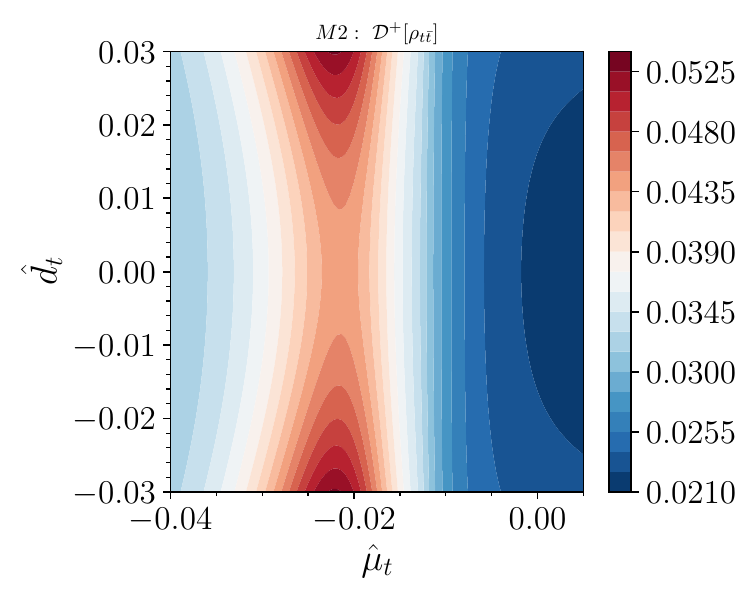}
    \includegraphics[width=0.32\textwidth]{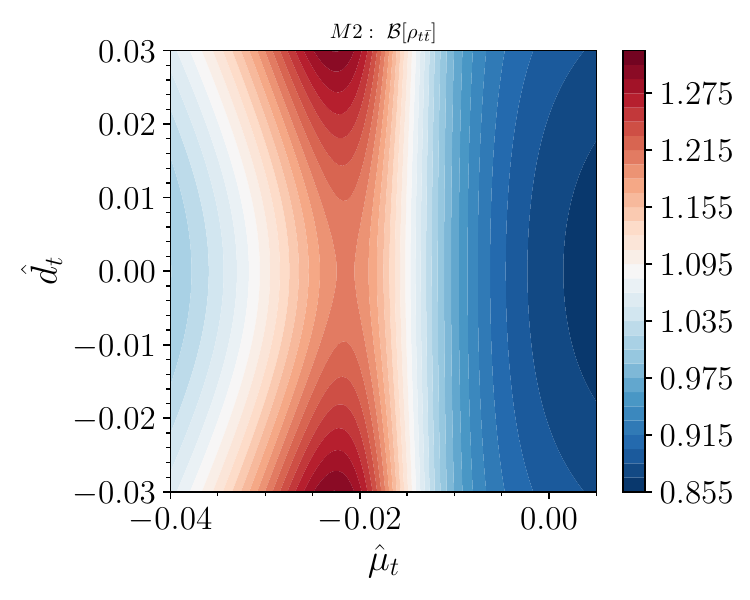}
    \caption{Contour of concurrence, geometric quantum discord and Bell variable as a function of CP-even and CP-odd anomalous chromo dipole moments. The distribution are shown for two binned $m_{t\bar{t}}$ viz. threshold M1, (top row) and intermediate region M2, (bottom row) at the parton level for $pp\to t\bar{t}\to b\bar{b}\ell^+\ell^-\nu_\ell\bar{\nu}_\ell$ at the $13$ TeV.}
    \label{fig:twod_quant}
\end{figure}
In Fig.~\ref{fig:twod_quant}, we show contour maps of $\mathcal{C}[\rho_{t\bar{t}}]$, 
$\mathcal{D}^+[\rho_{t\bar{t}}]$, and $\mathcal{B}[\rho_{t\bar{t}}]$ in the 
$(\hat{\mu}_t, \hat{d}_t)$ plane for the threshold bin M1 (top row) and the 
intermediate region M2 (bottom row), with both couplings restricted to their 
experimentally allowed ranges.

In the threshold region M1, the contour structure is dominated by the $\hat{\mu}_t$ 
dependence, with nearly vertical banding across the entire $\hat{d}_t$ range. All 
three observables peak sharply at $\hat{\mu}_t \approx -0.025$, independent of 
$\hat{d}_t$, with concurrence reaching $\sim 0.40$, GQD $\sim 0.16$, and Bell 
parameter $\sim 1.85$. The $\hat{d}_t$ dependence is almost entirely absent in this 
bin, consistent with the CP-odd operator entering CP-even observables only 
quadratically and its contribution being negligible within the narrow experimentally 
allowed range $|\hat{d}_t| \leq 0.03$. The Bell parameter spans $1.61$--$1.85$ 
across the entire plane, remaining well below the classical bound of $2$, confirming 
that Bell inequality violation does not occur in the threshold region within the 
experimentally allowed coupling space. The GQD remains non-zero throughout the 
entire plane, including regions where the concurrence is already suppressed near 
$\hat{\mu}_t \approx 0$, confirming the persistence of non-classical correlations 
beyond QE.

The intermediate bin M2 presents a qualitatively distinct and richer structure. The 
contour pattern transitions from the vertical banding of M1 to a clear X-shaped pattern, with two localized enhancements appearing at 
$\hat{\mu}_t \approx -0.025$ and $|\hat{d}_t| \approx 0.02$--$0.03$. The 
concurrence is nearly vanishing near the SM point $(\hat{\mu}_t, \hat{d}_t) \approx 
(0, 0)$ and along the $\hat{\mu}_t \approx 0$ axis, confirming that the spin state 
is effectively separable in the SM limit of this kinematic bin. The GQD remains 
finite over much of this same region, demonstrating that separability does not imply 
classicality: non-classical correlations persist below QE  threshold and 
are captured by QD. The Bell parameter in M2 ranges between $0.855$ and 
$1.275$, remaining well below $2$ across the entire experimentally allowed plane, 
with the lowest values near the SM point and the highest values at the wings of the 
$\hat{\mu}_t$ distribution.

The X-shaped pattern and the asymmetric tilt of the contours under 
$\hat{d}_t \to -\hat{d}_t$ in M2 are direct signatures of the CP-odd mixed 
interference term $\propto \hat{\mu}_t \hat{d}_t$ in the spin density matrix. This 
term is linear in $\hat{d}_t$ with a $\hat{\mu}_t$-dependent coefficient, 
schematically
\begin{align}
    \mathcal{O}_{\mathrm{CP\text{-}odd}} \;\sim\; a\,\hat{d}_t \;\pm\; 
    b\,\hat{\mu}_t\hat{d}_t,
\end{align}
which geometrically manifests as tilted, axis-asymmetric contours that break the 
$\hat{d}_t \to -\hat{d}_t$ reflection symmetry when $\hat{\mu}_t \neq 0$. This 
sign flip is clearly visible in all three contour maps of M2 and constitutes a 
direct imprint of CP-odd dynamics on the quantum information observables of the 
$t\bar{t}$ spin state, observable within the experimentally allowed coupling space.

\subsection{Non-classicality dependence  on anomalous weak dipole moments} 
Having discussed the impact of anomalous chromo-dipole interactions, we now turn to 
the effects of anomalous weak dipole moments of the top quark. These are 
parameterized by the operators $\Delta C_1^{A,V}$ and $C_2^{A,V}$, which modify the 
$t\bar{t}Z$ vertex and consequently alter the $q\bar{q} \to t\bar{t}$ production 
amplitude. Although the quark--annihilation channel is sub-dominant relative to 
gluon fusion at the LHC, anomalous weak dipole couplings can nonetheless deform the 
spin density matrix of the produced $t\bar{t}$ pair and thereby modify the quantum 
information observables accessible through the decay products. All four couplings are 
varied within the range $[-0.5, 0.5]$, consistent with the approximate 68\% CL 
constraints reported in Ref.~\cite{Machethe:2025oev}.

\begin{figure}[!htb]
    \centering
    \includegraphics[width=0.98\textwidth]{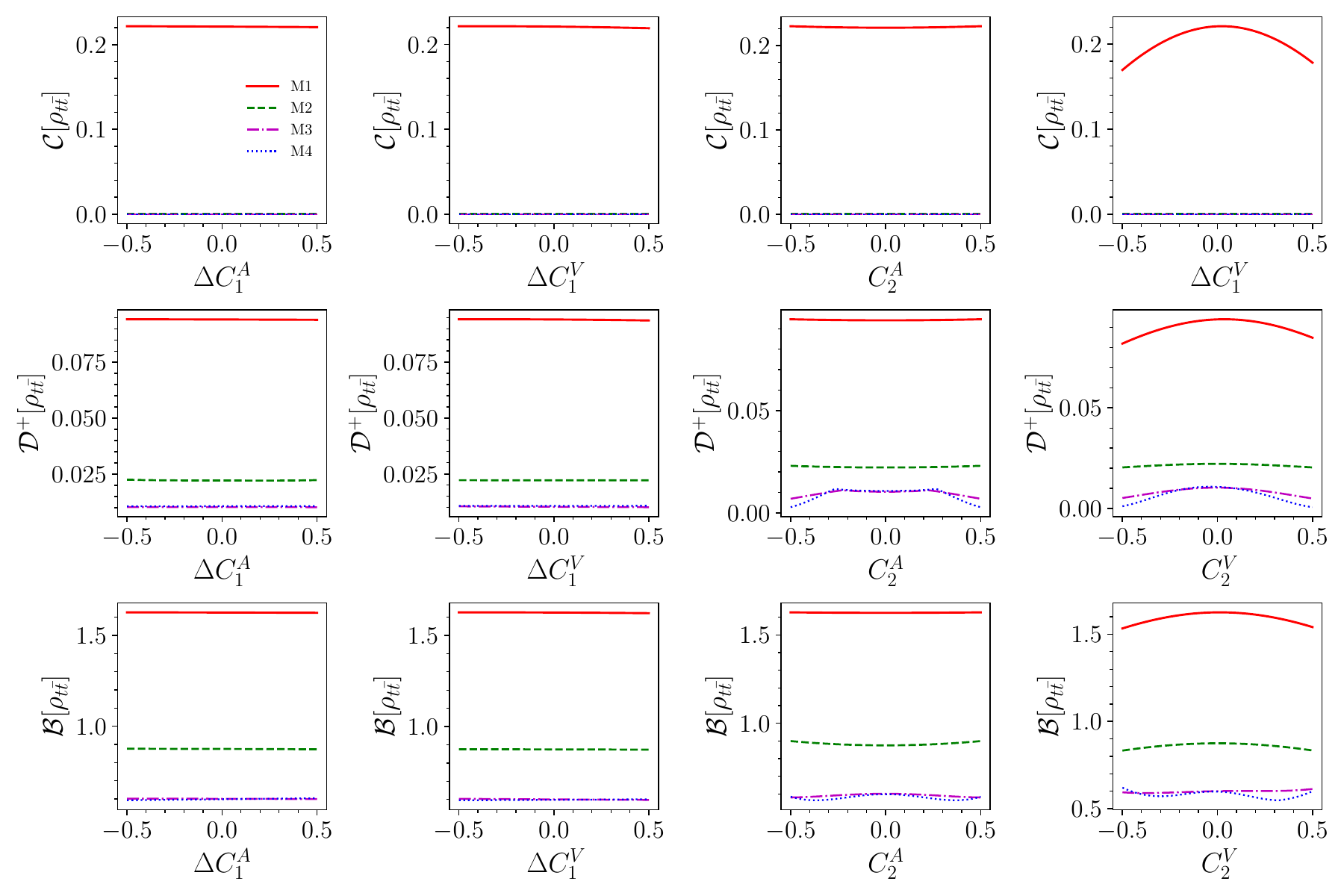}
    \caption{Behavior of concurrence, geometric discord and Bell variable w.r.t to anomalous weak dipole moments. The distribution are at the parton level for $pp\to t\bar{t}\to l^+l^-b\bar{b}\nu_l\bar{\nu}_l$ process at $\sqrt{s}=13$ TeV. }
    \label{fig:weak_quant}
\end{figure}

Figure~\ref{fig:weak_quant} displays the dependence of $\mathcal{C}[\rho_{t\bar{t}}]$, 
$\mathcal{D}^+[\rho_{t\bar{t}}]$, and $\mathcal{B}[\rho_{t\bar{t}}]$ on each of the 
four anomalous weak dipole coefficients across all four $m_{t\bar{t}}$ bins. The 
results divide cleanly into two qualitatively distinct regimes depending on the 
operator class.

For the vector and axial-vector current modifications $\Delta C_1^A$ and $\Delta C_1^V$ 
(first and second columns), all three observables are completely flat as a function 
of the coupling across the entire displayed range and for all bins. The concurrence 
is pinned at its SM value of $\sim 0.21$ for M1 and effectively zero for M2--M4; 
the GQD remains at $\sim 0.075$ (M1) and $\sim 0.025$ (M2), with M3 and M4 
nearly flat at $\lesssim 0.01$; and the Bell parameter holds steady at $\sim 1.65$, 
$\sim 0.85$, and $\sim 0.63$ for M1, M2, and M3/M4 respectively. This complete 
insensitivity reflects the negligible interference of these operators with the SM 
amplitude in shaping the spin-correlation structure of the $t\bar{t}$ system, 
consistent with their suppressed contribution to the $q\bar{q}$-initiated production 
channel at LHC energies.

A qualitatively different picture emerges for the dipole operators $C_2^A$ and 
$C_2^V$. For the CP-odd coupling $C_2^A$ (third column), the threshold bin M1 
remains fully insensitive, with all three observables flat at their SM values across 
the entire displayed range. In the higher invariant-mass bins, a mild but structured 
variation becomes visible. The GQD in M2 develops a shallow symmetric dip near 
$C_2^A = 0$ before recovering toward the wings, while M3 and M4 exhibit a slightly 
more pronounced suppression near the origin. The Bell parameter reflects the same 
pattern, with M2 showing a broad bowl-shaped minimum near $C_2^A = 0$ and M3/M4 
dipping more sharply to $\sim 0.6$--$0.7$ near the origin. The strict symmetry of 
all distributions about $C_2^A = 0$ confirms that this CP-odd operator enters 
CP-even observables only at quadratic order, consistent with the parametric structure 
of the spin density matrix. The concurrence remains negligibly small for M2--M4 
throughout the displayed range, indicating that the CP-odd weak dipole does not 
generate appreciable entanglement within the experimentally constrained coupling 
space.

The most significant sensitivity is observed for the CP-even coupling $C_2^V$ 
(fourth column). In the threshold bin M1, all three observables display a pronounced 
inverted-parabolic dependence, peaking near $C_2^V = 0$ with concurrence $\sim 0.21$, 
GQD $\sim 0.075$, and Bell parameter $\sim 1.65$, all falling monotonically toward 
the wings of the displayed range. The inverted-parabolic shape reflects the dominant 
quadratic dependence of the diagonal spin-correlation matrix elements on $C_2^V$. 
For M2, a similarly shaped but broader and attenuated inverted parabola is observed 
in the GQD and Bell parameter, while the concurrence remains negligibly small. For 
M3 and M4, the response is further suppressed, with a narrow central peak in the 
GQD and Bell parameter that falls steeply toward the wings, indicating that the 
sensitivity to $C_2^V$ becomes increasingly concentrated near the SM point at higher 
invariant masses. Across all bins and for all four weak dipole operators, the Bell 
parameter remains well below the classical bound of $2$, confirming that Bell 
inequality violation does not occur within the allowed weak dipole 
coupling space considered here.

\begin{figure}[!htb]
    \centering
    \includegraphics[width=0.32\textwidth]{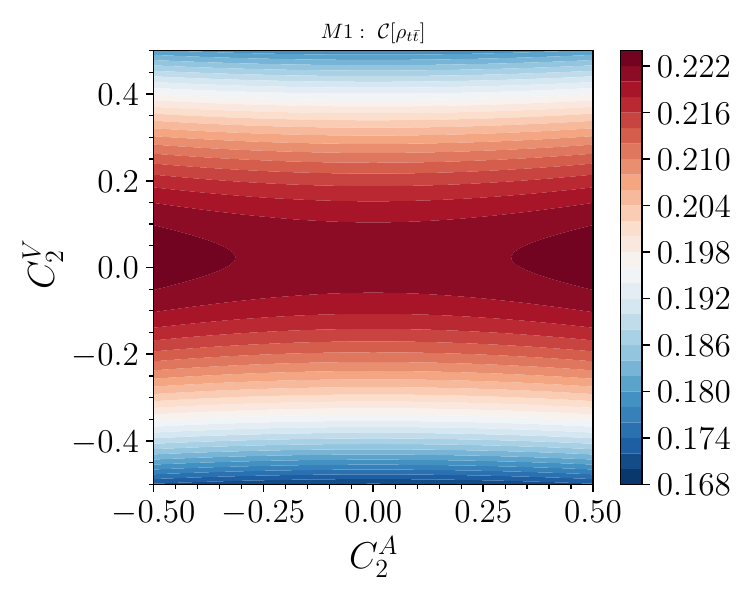}
    \includegraphics[width=0.32\textwidth]{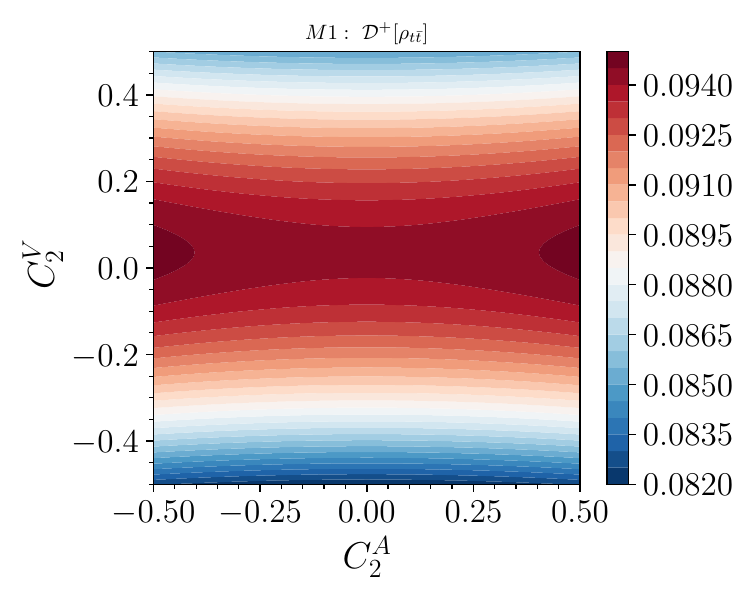}
    \includegraphics[width=0.32\textwidth]{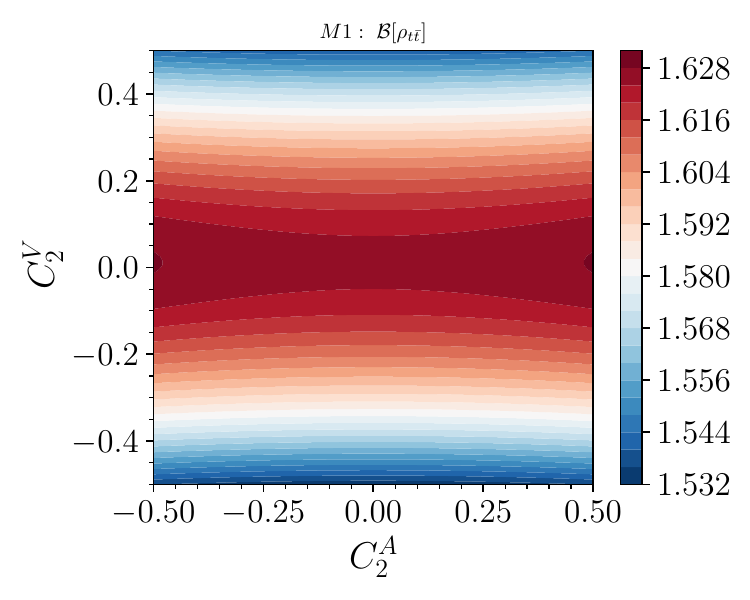}
    \includegraphics[width=0.32\textwidth]{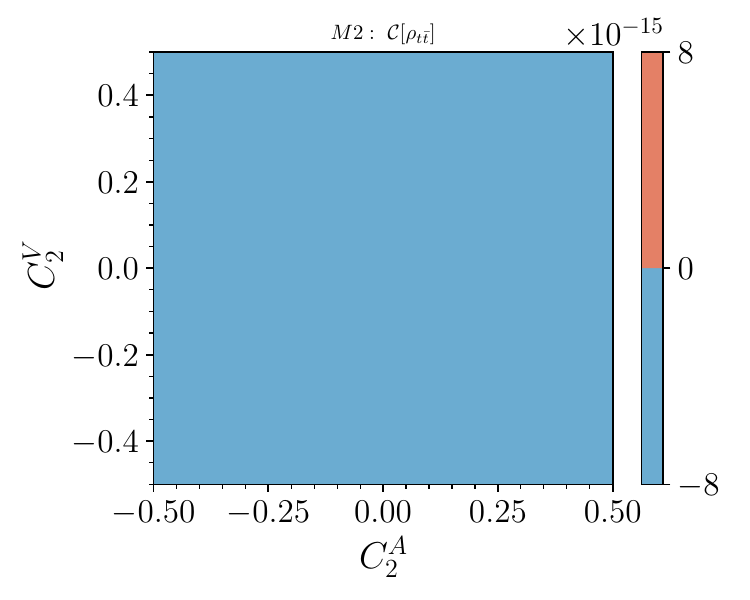}
    \includegraphics[width=0.32\textwidth]{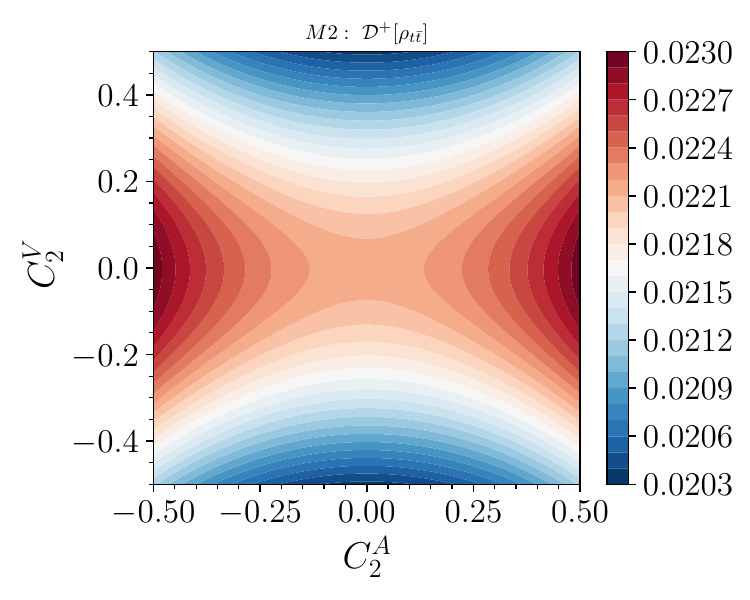}
    \includegraphics[width=0.32\textwidth]{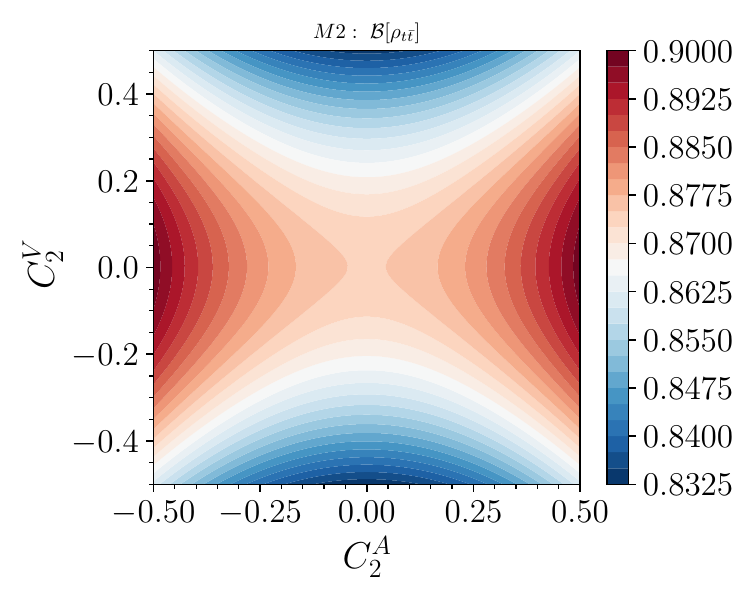}
    \caption{Contours of concurrence, GQD and Bell variable as a function of anomalous $CP$-odd $C_2^A$ and CP-even $C_2^V$ weak dipole coupling. The distribution are shown at the parton level for $pp\to t\bar{t}$ production process followed by leptonic decay at $\sqrt{s}=13$ TeV for two binned $m_{t\bar{t}}$ region, viz., i) Threshold region, M1 (top row) and ii) Intermediate region, M2 (bottom row).}
    \label{fig:2d_weak_quant}
\end{figure}

In Fig.~\ref{fig:2d_weak_quant}, we show the contour distributions of 
$\mathcal{C}[\rho_{t\bar{t}}]$, $\mathcal{D}^+[\rho_{t\bar{t}}]$, and 
$\mathcal{B}[\rho_{t\bar{t}}]$ in the $(C_2^A, C_2^V)$ parameter plane for the 
threshold region M1 (top row) and the intermediate region M2 (bottom row), with 
both couplings varied within the range $[-0.5, 0.5]$.

In the threshold region M1, all three observables are dominated by the $C_2^V$ 
dependence, exhibiting a clear horizontal banding structure across the entire 
$C_2^A$ range. The concurrence peaks at $\sim 0.222$ near $C_2^V \approx 0$ and 
decreases monotonically to $\sim 0.168$ at $|C_2^V| \approx 0.5$, while remaining 
essentially insensitive to $C_2^A$ throughout. The GQD and Bell parameter display 
the same pattern, with peak values of $\sim 0.094$ and $\sim 1.628$ respectively 
near $C_2^V \approx 0$, both falling gradually toward the $C_2^V$ wings. This 
behaviour is consistent with the one-dimensional scans and reflects the dominant 
role of the CP-even dipole operator in modifying the elements of the 
spin-correlation tensor through its quadratic dependence. The $C_2^A$ dependence 
is negligible in M1, as expected from the quadratic entry of the CP-odd operator 
into CP-even observables and its suppressed contribution within the narrow 
experimentally allowed range. The Bell parameter spans $1.532$--$1.628$ across 
the entire plane, remaining well below the classical bound of $2$.

The intermediate region M2 presents a qualitatively distinct and more intricate 
picture. Most strikingly, the concurrence is vanishingly small throughout the 
entire $(C_2^A, C_2^V)$ plane, with values at the level of numerical precision 
($\sim 10^{-15}$), confirming that the $t\bar{t}$ spin state is fully separable 
in this kinematic bin for all weak dipole coupling values within the displayed 
range. In contrast, the GQD and Bell parameter develop a clear four-lobe or 
X-shaped structure, with enhancements appearing symmetrically in all four 
quadrants of the $(C_2^A, C_2^V)$ plane and suppression along both axes. The 
GQD reaches $\sim 0.023$ at the off-axis maxima and falls to $\sim 0.0203$ near 
the origin and along the coordinate axes, while the Bell parameter peaks at 
$\sim 0.900$ in the off-axis lobes and dips to $\sim 0.833$ near the origin. The 
strict symmetry of the pattern under independent sign reversals $C_2^A \to -C_2^A$ 
and $C_2^V \to -C_2^V$ confirms that dominant contribution of both operators enter the relevant spin density 
matrix elements at even order in M2. The four-lobe structure signals a genuine 
two-dimensional interplay between $C_2^A$ and $C_2^V$: the enhancements occur 
where both couplings are simultaneously non-zero, reflecting a mixed quadratic 
contribution $\propto (C_2^A)^2 (C_2^V)^2$ or cross-term structure in the 
spin-correlation matrix. The GQD remaining non-zero and structured throughout the 
plane, despite the complete absence of entanglement, again demonstrates that 
separability does not imply classicality and that GQD provides 
a more sensitive probe of the non-classical structure of the $t\bar{t}$ spin state 
than concurrence in this kinematic regime. The Bell parameter remains well below $2$ 
across the entire plane, implying the correlations are purely local within 
the weak dipole coupling space.
\begin{figure}[!t]
    \centering
    \includegraphics[width=0.49\linewidth]{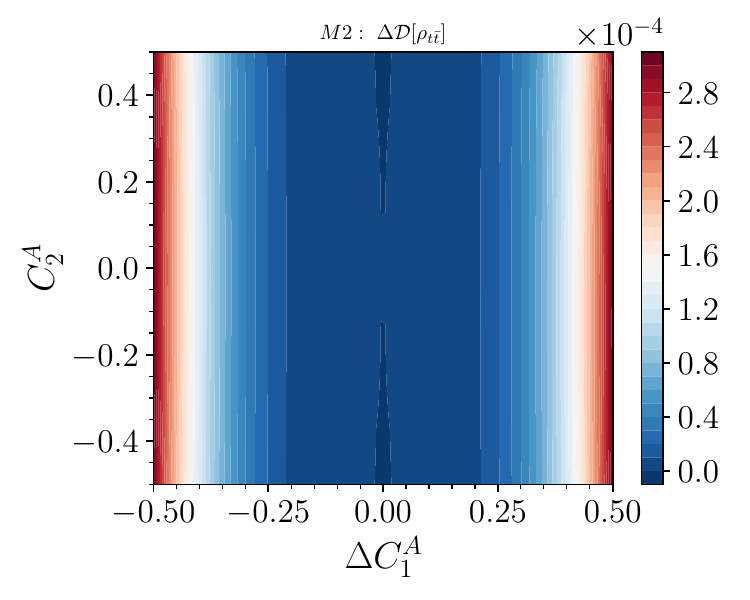}
     \includegraphics[width=0.49\linewidth]{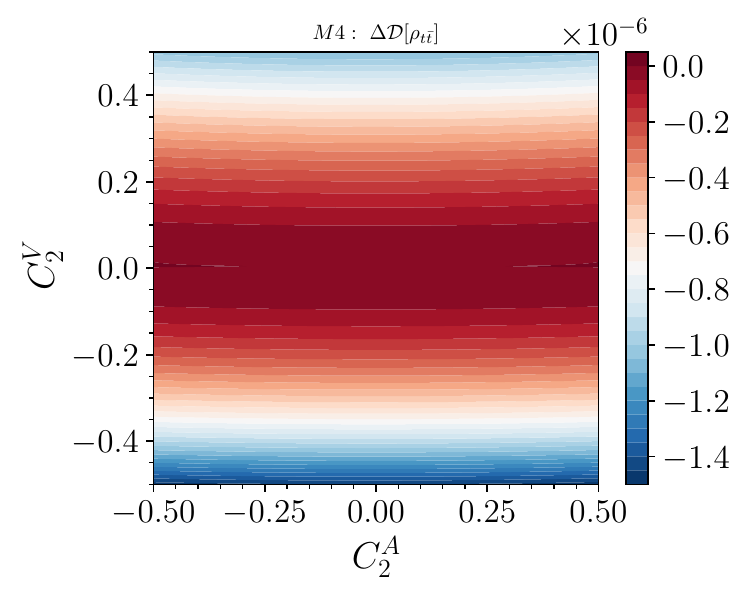}
    \caption{Distribution of difference on the GQD ($\Delta \mathcal{D}[\rho_{t\bar{t}}]$) as a function of two anomalous weak dipole moments. The distribution are shown at the parton level for $pp\to t\bar{t} \to b\bar{b}\ell^+\ell^-\nu_\ell\bar{\nu}_\ell$ at $\sqrt{s}=13$ TeV.}
    \label{fig:dgqd}
\end{figure}

The difference of GQD, $\Delta\mathcal{D}[\rho_{t\bar{t}}] = 
\mathcal{D}^+[\rho_{t\bar{t}}] - \mathcal{D}^-[\rho_{t\bar{t}}]$, provides a 
direct probe of CP-violating effects in the $t\bar{t}$ spin state. The superscript 
$+$ ($-$) denotes that the projective measurement is performed on the top (anti-top) 
quark subsystem, with the discord quantifying the residual quantum correlations in 
the complementary subsystem. Under CP symmetry, the two measurements are related by 
the transformation $t \leftrightarrow \bar{t}$, so any non-vanishing 
$\Delta\mathcal{D}$ signals CP violation.

Figure~\ref{fig:dgqd} displays $\Delta\mathcal{D}[\rho_{t\bar{t}}]$ as a function 
of two pairs of anomalous weak dipole couplings at the parton level for 
$pp \to t\bar{t} \to b\bar{b}\ell^+\ell^-\nu_\ell\bar{\nu}_\ell$ at 
$\sqrt{s} = 13~\mathrm{TeV}$, with all couplings varied within the range $[-0.5, 0.5]$. In the intermediate region M2 (left panel), 
$\Delta\mathcal{D}$ is shown in the $(\Delta C_1^A, C_2^A)$ plane. The distribution 
exhibits a clear vertical banding structure, with $\Delta\mathcal{D}$ vanishing 
along the $\Delta C_1^A = 0$ axis and growing monotonically as $|\Delta C_1^A|$ 
increases, reaching values of order $\sim 2.8\times 10^{-4}$ at the edges of the 
displayed range. The distribution is strictly non-negative and symmetric under 
$C_2^A \to -C_2^A$, reflecting the even-order entry of the CP-odd coupling $C_2^A$ 
into the spin density matrix. The dominant sensitivity thus arises entirely from 
$\Delta C_1^A$, which breaks the $t \leftrightarrow \bar{t}$ symmetry linearly and 
generates a non-zero discord asymmetry proportional to its magnitude.

In the extreme boosted region M4 (right panel), $\Delta\mathcal{D}$ is shown in 
the $(C_2^A, C_2^V)$ plane. The distribution is strictly non-positive across the 
entire plane, ranging from $0$ near $C_2^V \approx 0$ to $\sim -1.4\times 10^{-6}$ 
at $|C_2^V| \approx 0.4$--$0.5$, with nearly no dependence on $C_2^A$. The 
horizontal banding confirms that the CP-odd asymmetry in M4 is driven primarily by 
$C_2^V$ rather than $C_2^A$, and that the response is quadratic in $C_2^V$ — 
consistent with the indirect generation of a CP-odd asymmetry through the interplay 
of $C_2^V$ with the asymmetric kinematic structure of the boosted regime. The 
magnitude is suppressed by two orders relative to M2, at the level of 
$\sim 10^{-6}$.

Taken together, these results confirm that $\Delta\mathcal{D}[\rho_{t\bar{t}}]$ is 
sensitive to CP-violating operator structures and responds differently to different 
coupling combinations across kinematic regions. We note, however, that the magnitudes 
observed — $\mathcal{O}(10^{-4})$ in M2 and $\mathcal{O}(10^{-6})$ in M4 — are 
extremely small, posing a severe challenge for any near-term experimental 
measurement. Given the precision achievable in current LHC analyses of $t\bar{t}$ 
spin correlations, which is typically at the level of $\mathcal{O}(10^{-2})$ or 
larger, these effects lie well below the reach of present experiments and would 
require a substantial improvement in both statistical precision and systematic 
control to be observed. The results therefore serve primarily as a theoretical 
characterization of the CP-odd imprint on quantum information observables within 
the experimentally allowed coupling space, rather than as immediate experimental 
targets.

\section{Conclusion and Discussion}
\label{sec:con}
In this paper, we have investigated the QI  quantities of the 
$t\bar{t}$ system produced in $pp$ collisions at $\sqrt{s} = 13$ TeV, focusing on 
three complementary QI quantities:  concurrence 
$\mathcal{C}[\rho_{t\bar{t}}]$, which is a measure of QE, GQD 
$\mathcal{D}^\pm[\rho_{t\bar{t}}]$, and the Bell parameter 
$\mathcal{B}[\rho_{t\bar{t}}]$. The spin density matrix of the $t\bar{t}$ pair is 
reconstructed from the joint angular distribution of the charged decay leptons in 
the $k$-$r$-$n$ helicity basis, and all results are reported at the parton level 
with leptonic top-pair decays. To systematically probe deviations from the SM, we 
adopt the SMEFT framework and consider a set of dimension-6 operators modifying the 
$gt\bar{t}$ and $t\bar{t}Z$ vertices, parameterized by six anomalous couplings 
contributing to the chromo- and weak dipole moments of the top quark. Crucially, 
all anomalous couplings are varied strictly within their current experimentally 
allowed ranges throughout this analysis.

Within the SM, the Bloch vectors of the top and anti-top quarks vanish identically 
due to the discrete symmetries of the production amplitude. QE  is 
present only in the threshold region $m_{t\bar{t}} \lesssim 400$ GeV, where the 
spin-singlet component of the $t\bar{t}$ state dominates. In the higher 
invariant-mass bins, $\mathcal{C}[\rho_{t\bar{t}}]$ is effectively zero, yet the 
GQD remains non-vanishing across the entire phase space, 
demonstrating that quantum correlations persist even in kinematic regions where the 
state is fully separable. This confirms the well-known hierarchy: QE 
implies QD, but the converse does not hold. The Bell parameter remains below 
the classical bound of $2$ throughout the SM invariant-mass spectrum, indicating 
the absence of genuine non-locality in the SM $t\bar{t}$ spin state for the 
configurations considered.

In the presence of anomalous chromo-dipole moments, the CP-even coupling $\hat{\mu}_t$ 
and CP-odd coupling $\hat{d}_t$ produce qualitatively distinct signatures, with 
both varied within their CMS-constrained ranges $\hat{\mu}_t \in [-0.04, +0.005]$ 
and $|\hat{d}_t| \leq 0.03$. For $\hat{\mu}_t$, all three observables exhibit a 
pronounced asymmetric peak at $\hat{\mu}_t \approx -0.025$, driven by constructive 
linear interference of the chromo-magnetic dipole operator with the SM amplitude. 
Notably, this peak coincides with the CMS central value of the chromo-magnetic 
dipole moment, implying that the current best-fit value would predict an enhancement 
of quantum correlations in the threshold bin relative to the pure SM prediction. The 
sensitivity is overwhelmingly concentrated in M1, with M2--M4 showing negligible 
variation. For the CP-odd coupling $\hat{d}_t$, all distributions are strictly 
symmetric about $\hat{d}_t = 0$, as required by the even-order entry of CP-odd 
operators into CP-even observables. Within the experimentally allowed range, the 
response is mild across all bins, with no Bell inequality violation observed for 
either coupling. 

For the anomalous weak dipole moments, all four couplings are varied within the 
range $[-0.5, 0.5]$, consistent with existing experimental constraints. The current 
modifications $\Delta C_1^{A,V}$ leave all three QI quantities entirely 
insensitive across all bins, reflecting their negligible contribution to the 
spin-correlation structure via the sub-dominant $q\bar{q}$-initiated production 
channel at LHC energies. The CP-odd dipole $C_2^A$ induces a mild symmetric 
suppression of the discord and Bell parameter near $C_2^A = 0$ in M2--M4, while 
leaving M1 unaffected. The most significant sensitivity is found for the CP-even 
coupling $C_2^V$, which produces a pronounced inverted-parabolic dependence in M1 
for all three quantities. In the two-dimensional $(C_2^A, C_2^V)$ plane, M1 is 
dominated by horizontal banding driven by $C_2^V$, while M2 develops a four-lobe 
structure in the GQD and Bell parameter reflecting a genuine two-dimensional 
interplay between both couplings — with the concurrence vanishing identically 
throughout M2, confirming complete separability. No Bell inequality violation is 
observed anywhere within the experimentally allowed weak dipole coupling space.

The difference of GQD, $\Delta\mathcal{D}[\rho_{t\bar{t}}]$, 
serves as a theoretically well-motivated probe of CP violation, vanishing 
identically when CP is conserved. Within the experimentally allowed coupling space, 
$\Delta\mathcal{D}$ reaches $\mathcal{O}(10^{-4})$ in M2 for the 
$(\Delta C_1^A, C_2^A)$ plane and $\mathcal{O}(10^{-6})$ in M4 for the 
$(C_2^A, C_2^V)$ plane. While these signals correctly encode the CP-odd structure 
of the underlying operator combinations, their magnitudes lie well below the 
precision achievable in current LHC analyses of $t\bar{t}$ spin correlations, 
which is typically at the level of $\mathcal{O}(10^{-2})$ or larger. 
$\Delta\mathcal{D}$ therefore currently serves as a theoretical characterization 
of the CP-odd imprint on quantum information observables rather than an immediate 
experimental target, and would require a substantial improvement in both statistical 
and systematic precision to be observed at the LHC.

The present study is necessarily limited in scope. The analysis is restricted to a 
subset of dimension-6 SMEFT operators at the parton level, and several important 
directions remain to be pursued. A comprehensive treatment incorporating the full 
basis of dimension-6 operators, along with dimension-8 contributions that become 
increasingly important in the boosted regime, would provide a more complete picture 
of the EFT sensitivity of quantum information observables. A realistic detector-level 
study including parton showering, hadronization, and experimental resolution effects 
is essential before direct comparison with LHC data. Furthermore, dedicated 
investigation of the interplay between QI quantities and 
conventional CP-violation searches is warranted, given the complementary sensitivity 
offered by $\Delta\mathcal{D}[\rho_{t\bar{t}}]$ to CP-odd top-quark interactions.

\section*{Acknowledgements}violat
This work was supported by  National Natural Science Foundation of China under Grant Nos. T2241005 and 12075059, as well as the startup fund of USTC.

\bibliographystyle{JHEP} 
\bibliography{refer.bib}
\end{document}